\documentclass[twocolumn,showpacs,preprintnumbers,amsmath,amssymb]{revtex4-1}
\usepackage{textcomp}

\usepackage{graphicx}
\begin{document}

\title{Direct measurement of DNA-mediated adhesion between lipid bilayers}
\author{S. F. Shimobayashi}
\affiliation{Department of Physics, Graduate School of Science, Kyoto University, Kyoto 606-8502, Japan}
\author{B. M. Mognetti}
\affiliation{Interdisciplinary Center for Nonlinear Phenomena and Complex Systems \&
Service de Physique des Syst\'{e}mes Complexes et M\'{e}canique Statistique, Universit\'{e} libre de Bruxelles, Campus Plaine, CP 231, Blvd du Triomphe, B-1050 Brussels, Belgium.}
\author{L. Parolini}
\affiliation{Biological and Soft Systems Sector, Cavendish Laboratory, University of Cambridge, CB3 0HE, Cambridge, United Kingdom}
\author{D. Orsi}
\affiliation{Department of Physics and Earth Sciences, University of Parma, I-43124, Parma, Italy}
\author{P. Cicuta}
\thanks{Email address}
\email{pc245@cam.ac.uk}
\author{L. Di Michele}
\thanks{Email address}
\email{ld389@cam.ac.uk}
\affiliation{Biological and Soft Systems Sector, Cavendish Laboratory, University of Cambridge, CB3 0HE, Cambridge, United Kingdom}
\date{\today}

\begin{abstract}
Multivalent interactions between deformable mesoscopic units are ubiquitous in biology, where membrane macromolecules mediate the interactions between neighbouring living cells and between cells and solid substrates. Lately, analogous artificial materials have been synthesised by functionalising the outer surface of compliant Brownian units, for example emulsion droplets and lipid vesicles, with selective linkers, in particular short DNA sequences. This development extended the range of applicability of DNA as a selective glue, originally applied to solid nano and colloidal particles. On very deformable lipid vesicles, the coupling between statistical effects of multivalent interactions and mechanical deformation of the membranes gives rise to complex emergent behaviours, as we recently contributed to demonstrate [Parolini \textit{et al., Nature Communications}, 2015, \textbf{6}, 5948]. Several aspects of the complex phenomenology observed in these systems still lack a quantitative experimental characterisation and fundamental understanding. Here we focus on the DNA-mediated multivalent interactions of a single liposome adhering to a flat supported bilayer. This simplified geometry enables the estimate of the membrane tension induced by  the DNA-mediated adhesive forces acting on the liposome. Our experimental investigation is completed by morphological measurements and the characterisation of the DNA-melting transition, probed by in-situ F\"{o}rster Resonant Energy Transfer spectroscopy. Experimental results are compared with the predictions of an analytical theory that couples the deformation of the vesicle to a full description of the statistical mechanics of mobile linkers. With at most one fitting parameter, our theory is capable of semi-quantitatively matching experimental data, confirming the quality of the underlying assumptions.
\end{abstract}
\maketitle



\begin{figure}[ht!]
\centering
  \includegraphics[width=8.8cm]{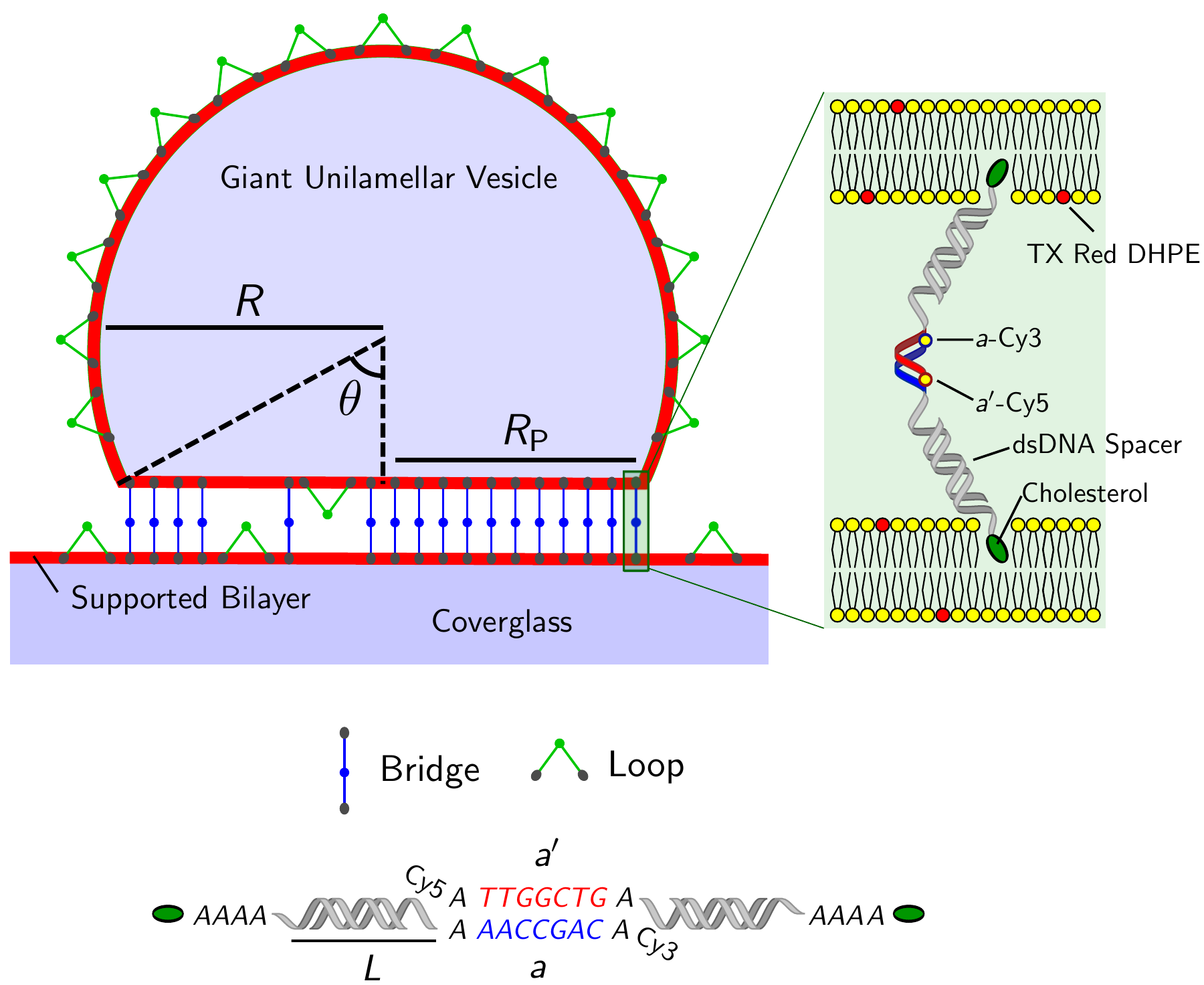}
  \caption{Not-to-scale schematics of a DNA-functionalised GUV adhering to a supported bilayer and details of the functionalisation.}
  \label{Figure1}
\end{figure}

Due to both fundamental and practical interest, the complex phenomenology of multivalent selective interactions has recently been a hot research topic in Soft Matter and Physical Chemistry. One of the main driving forces behind this effort has been the development of self-assembly strategies based on DNA-mediated multivalent interactions.\cite{Di-Michele_PCCP_2013}Introduced by the seminal works of Mirkin\cite{Mirkin_Nature_1996} and Alivisatos,\cite{Alivisatos_Nature_1996} who nearly two decades ago demonstrated the selective DNA-mediated self-assembly of gold nanoparticles, this approach has been optimised to the point of mastering the structure of multicomponent crystal lattices\cite{Hill_NanoLetters_2008,Maye_NatNano_2010,Nykypanchuk_Nature_2008,Park_NMat_2010,Mirkin_Science_2011} and amorphous materials.\cite{Di-Michele_NatComm_2013,Varrato_PNAS_2012,Di-Michele_SoftMatter_2014} Applications of these self-assembly strategies span from photonics and plasmonics,\cite{Young_AdvMat_2014} to biosensing\cite{Thaxton_PNAS_2009}, and gene therapy.\cite{Wu_JACS_2014,Rosi_Science_2006,Cutler_JACS_2012} An abundance of experimental studies called for the development of models to understand the complex statistical mechanics of DNA-mediated multivalent interactions between solid particles.\cite{Dreyfus_PRL_2009,Varilly_JChemPhys_2012,Angioletti-Uberti_JCP_2013,Mognetti_SoftMatter_2012,Mognetti_PNAS_2012,Leunissen_JCP_2011} More recently, the same self-assembly strategy has been applied to compliant Brownian units, including emulsion droplets,\cite{Hadorn_PNAS_2012,Feng_SoftMatter_2013} and lipid vesicles.\cite{Hadorn_Langmuir_2013,Hadorn_PLOSOne_2010,Beales_JPCA_2007,Beales_SoftMatter_2011,Beales_ACIS_2014} Especially for the case of liposomes, which are significantly more deformable than emulsion droplets, DNA-mediated adhesion plays at a similar magnitude as other forces, leading to rich coupling.\cite{Beales_SoftMatter_2011,Parolini_NatComms_2015} Moreover, on these liquid surfaces, DNA linkers can freely diffuse, and the confinement induced by binding results in significant entropic contributions to the hybridisation free energy.\\
We recently demonstrated that the coupling between mechanical deformability of Giant-Unilamellar-Veisicles (GUVs) and DNA-mediated interactions, results in unexpected emergent response to temperature changes, leading to negative thermal expansion and tuneable porosity of DNA-GUV assemblies.\cite{Parolini_NatComms_2015} By means of an analytical theory, that jointly describes the geometrical features of the GUVs and the statistical mechanics of DNA tethers,\cite{Parolini_NatComms_2015,Angioletti-Uberti_PRL_2014} we rationalised experimental evidence and highlighted the importance of translational entropy in coupling the DNA-binding free energy to the morphology of the substrates.
Nonetheless, a further experimental investigation is necessary to assess all the aspects of the complex phenomenology of DNA-GUV conjugates, and test the accuracy of the theoretical framework applied to their description. In particular, a direct measurement of the DNA-mediated adhesive forces and the quantitative characterisation of the melting transition would represent new strong tests of the current understanding.\\
 In this article, we present experiments aimed at measuring the temperature-dependence of the membrane tension induced on giant liposomes by DNA-mediated adhesion. Membrane tension is accessed by \emph{flickering} measurements, in which we reconstruct the spectrum of thermal fluctuations of the liposomes at the equatorial cross-sections.\cite{Yoon_BJ_2009,Helfrich,Pecreaux_EPJE_2004} This is a classic technique widely used to study vesicle tension and bending rigidity. To make these measurements possible, we adopt an experimental geometry in which liposomes do not interact with each other, instead they adhere to Supported Lipid Bilayers (SBLs) fabricated on rigid glass substrates, as demonstrated in Fig.~\ref{Figure1}. This allows imaging of the equator of the vesicle through confocal microscopy. Various morphological observables are also precisely assessed via 3-dimensional confocal imaging. Finally, the use of suitably labelled DNA tethers enables the direct assessment of the relative number of formed DNA bonds by means of in-situ F\"{o}rster Resonant Energy Transfer (FRET) spectroscopy. Experimental results are compared to predictions from our analytical theory, extended from ref.\cite{Parolini_NatComms_2015} to apply to this new geometry. Using at most one fitting parameter, we find semi-quantitative agreement for all the measured quantities, confirming the accuracy of the underlying assumptions of our model.\\
The remainder of this article is structured as follows. First we describe the experimental setup, including sample preparations protocols, materials, imaging and image analysis techniques. Then we outline our theoretical model focusing on the novel aspects introduced here for the GUV-plane geometry. Finally we discuss the experimental results and compare them with theoretical predictions.\\


\section{Experimental}
Figure~\ref{Figure1} shows a schematic of a DNA-functionalised GUV adhering to a DNA-functionalised SBL. GUVs and SBLs are prepared and separately functionalised with cholesterol labeled DNA constructs. The hydrophobic cholesterol inserts into the lipid bilayer allowing DNA tethers to freely diffuse. Connected to the cholesterol anchor there is a section of length $L=9.8$ nm (29 base pairs\cite{Smith_Science_1996}) of double-strands DNA (dsDNA), terminating in a single-stranded DNA (ssDNA) \emph{sticky end}, which mediates the attractive interactions. To further facilitate the pivoting motion of the DNA tethers, 4 unpaired adenine bases are left between the cholesterol anchor and the dsDNA spacer. One unpaired adenine is left between the dsDNA spacer and the sticky end. Two mutually complementary sticky ends are used: $a$ and $a'$, of 7 bases each (see Fig.~\ref{Figure1}). GUVs and SBLs are functionalised with equal molar fractions of both $a$ and $a'$. Tethers can therefore form intra-membrane \emph{loops} and inter-membrane \emph{bridges}. The latter are responsible for the observed adhesion and are confined within the contact area between the GUVs and the SBLs. We use FRET spectroscopy to estimate the fraction of formed DNA bonds. To enable these measurements the termini of the sticky ends $a$ and $a'$ are functionalised with a Cy3 (indocarbocyanine) and a Cy5 (indodicarbocyanine) molecules respectively, spaced by an unpaired adenine base. Note that the sticky ends used here are shorter than those adopted in our previous study\cite{Parolini_NatComms_2015} (7 base-pairs instead of 9), this choice was made to lower the melting temperature of the DNA to within an experimentally accessible temperature, and reduce the overall strength of the DNA-mediated adhesion, making it easier to measure by flickering spectroscopy.

\subsection{Materials and sample preparation}

\subsubsection*{GUVs electroformation~~}1,2-Dioleoyl-sn-glycero-3-phosphocholine (DOPC, Avanti Polar Lipids) GUVs are prepared by standard electroformation in 300 mM sucrose (Sigma Aldrich) solution in double-distilled water.\cite{Angelova_FD_1989,Angelova_PCPS_1992} Full details can be found in ref.\cite{Parolini_NatComms_2015}

\subsubsection{Supported bilayers}DOPC supported bilayers (SBLs) are formed by rupture of Small Unilamellar Vesicles (SUVs $\sim$ 100 nm) on the hydrophilised glass bottom of sample chambers.\cite{Cremer_JPCB_1999,Brian_PNAS_1984}
SUVs are prepared by standard extrusion. Briefly, 200 $\mu$l of 25 mg/ml DOPC solution in chloroform are left to dry in a glass vial, then hydrated by adding 500 $\mu$l of 300 mM sucrose solution and mixed by vortexing for at least 5 minutes. The solutions are then transferred in plastic vials and treated with 5 cycles of rapid freezing/unfreezing by alternatively immersing the vial in baths of liquid nitrogen and warm water.\cite{Beales_JPCA_2007} Extrusion is carried out using a hand-driven mini-extruder (Avanti Polar Lipids) with a polycarbonate track-etched membrane (100 nm pores, Whatman). To facilitate rupture of the SUVs and bilayer formation, the extruded solution is diluted in a 1:9 ratio in iso-osmolar solution containing TE buffer (10 mM tris(hydroxymethyl) aminomethane, 1~mM ethylenediaminetetra acetic acid, Sigma Aldrich), 5~mM MgCl$_{2}$ and 272~mM glucose (Sigma Aldrich).\\
Sample chambers are obtained by applying adhesive silicone-rubber multi-well plates (6.5 mm $\times$ 6.5 mm $\times$ 3.2 mm, Flexwell, Grace Biolabs) on glass coverslips (26$\times$ 60 mm, No.1, Menzel-Gls\"{a}er), cleaned following a previously reported protocol.\cite{Di-Michele_NatComm_2013} To form SBLs, the glass bottom of the cells is hydrophilised by plasma cleaning on a plasmochemical reactor (Femto, Diener electronic, Germany), operated at frequency of 40 kHz, pressure of 30 Pa, and power input of 100 W for 5 minutes. Within 5 minutes from plasma cleaning, each cell is filled with $100$ $\mu$l of diluted SUV solution and incubated for at least 30 minutes at room temperature to allow bilayer formation. To remove excess lipid and magnesium, the cells are repeatedly rinsed with the experimental solution (TE buffer, 100 mM NaCl, 87 mM glucose). Care is taken to always keep the bilayers covered in buffer to avoid exposure to air.\\
For experiments not involving FRET spectroscopy GUVs and SBLs are stained with 0.8-1.2\% molar fraction of Texas Red 1,2-Dihexadecanoyl-sn-Glycero-3-Phosphoethanolamine, Triethylammonium Salt (Texas Red DHPE, Life Technologies).

\subsubsection*{DNA preparation~~}
The DNA tethers are pre-assembled from two ssDNA strands, one of them (\emph{i}) functionalized with a cholesterol molecule and the second (\emph{ii}) carrying the sticky end:\\
\begin{itemize}
\item[\emph{i}] $5'$--{\bf CGT GCG CTG GCG TCT GAA AGT CGA TTG CG} {\it AAAA} [Cholesterol TEG] --$3'$
\item[\emph{ii}] $5'$--{\bf GC GAA TCG ACT TTC AGA CGC CAG CGC ACG} {\it A} [Sticky End] {\it A}  Cy3/Cy5 --$3'$.
\end{itemize}
The bold letters indicate the segments forming the dsDNA spacer, the italic letters the inert flexible spacers. DNA is purchased lyophilized from Integrated DNA Technologies, reconstituted in TE buffer, aliquoted, and stored at -20 $^\circ$C. For assembling the constructs, we dilute equal amounts of the two single strands, \emph{i} and \emph{ii}, to 1.6 $\mu$M in TE buffer containing 100 mM NaCl. Hybridization is carried out by ramping down the temperature from 90$^\circ$C to 20$^\circ$C at a rate of -0.2$^\circ$C min$^{-1}$ on a PCR machine (Eppendorf Mastercycler).\cite{Parolini_NatComms_2015}

\subsubsection*{Membrane functionalisation.~~}
Fuctionalisation of the supported bilayers is carried out by injecting 90 $\mu$L of iso-osmolar experimental solution (TE buffer, 87 mM glucose, 100 mM NaCl) containing 
overall $X$ moles of DNA constructs into each of the silicone-rubber cells, with equal molarity of $a$ and $a'$ strands. Similarly, GUVs are functionalised by diluting 10 $\mu$L of electroformed vesicle solution in 90 $\mu$L of iso-osmolar experimental solution containing $X$ moles of DNA constructs. GUVs and SLBs are incubated for at least 1 hour at room temperature to allow grafting.  After incubation, 10 $\mu$l of the the liposome solution are injected into the sample chambers, which are immediately closed with a second clean coverslip and sealed using rapid epoxy glue (Araldite). Care is taken to prevent the formation of air bubbles. Sedimentation of the GUVs results in the formation of an adhesion patch between GUVs and supported bilayer. In a typical sample a fraction of the GUVs is found to form clusters. We limit our analysis to isolated GUVs.\\
We tested samples at different DNA concentrations, obtained by setting $X$ to $X_\mathrm{low}=0.09,$ $X_\mathrm{med}=0.9$ and $X_\mathrm{high}=2.7$~pmoles. We have previously quantified the surface coverage of GUVs functionalised with $X_\mathrm{med}=0.9$ pmoles in $\rho_\mathrm{DNA}^\mathrm{med}=390\pm90$ $\mu$m$^{-2}$.\cite{Parolini_NatComms_2015} Here we proportionally assume $\rho_\mathrm{DNA}^\mathrm{low}=39\pm9$ $\mu$m$^{-2}$, $\rho_\mathrm{DNA}^\mathrm{high}=1200\pm300$ $\mu$m$^{-2}$. Having estimated that the overall surface of the SBL is approximately equal to the surface of the GUVs used for each sample, we assume equal DNA coverages for the SBL. This assumption is confirmed by fluorescence emission measurements: at sufficiently high temperature, when no DNA bridges are formed between GUVs and SBL, DNA is uniformly distributed on both interfaces. In this regime the fluorescence emission from DNA located within the contact area between a GUV and the SBL approximately equals twice the intensity measured on the free SBL ($2.1\pm 0.1$), confirming that GUVs and SBL have, within experimental errors, the same DNA coverage.\\

\subsection{Imaging and image analysis}

\subsubsection*{Imaging and temperature cycling.~~}
The samples are imaged on a Leica TCS SP5 laser-scanning confocal microscope. Texas Red DHPE is excited with a He-Ne laser (594 nm). For FRET spectroscopy measurements, Cy3 is excited with an Ar-ion laser line (514 nm). The temperature of the sample is regulated with a home-made Peltier device controlled by a PID (proportional-integral-derivative) controller, featuring a copper plate to which the sample chamber is kept in thermal contact. Two thermocouples are used as temperature sensors. The first sensor, kept in contact with the copper plate, serves as a feedback probe for the PID controller. The second thermocouple is inserted in a dummy experimental chamber, filled with water, and used to precisely probe the temperature of the sample. For all the temperature-dependent experiments imaging is carried out using a Leica HCX PL APO CS 40$\times$ 0.85 NA dry objective, to prevent heat dissipation.\\   
The temperature-dependent morphology of adhering GUVs is captured via confocal z-stacks and reconstructed using a custom script written in Matlab. Briefly: Each z-stack contains a single adhering GUV. The plane of the SBL/adhesion patch is identified as the one with maximum average intensity, then the adhesion patch is reconstructed by thresholding, following the application of a bandpass filter to flatten the background and remove pixel-level noise. From the area $A_\mathrm{p}$ of the adhesion patch we extract the patch-radius as $R_\mathrm{p}=\sqrt{A_\mathrm{p}/\pi}$. The portion of the z-stack above the SBL is then scanned, and in each plane the contour of the vesicle, suitably highlighted by filtering and thresholding, is fitted with a circle. The slice featuring the largest circle is identified as the equatorial plane, determining the vesicle radius $R$. The contact angle is derived as $\theta=\sin^{-1}{\left(R_\mathrm{p}/R\right)}$ (see Fig.~\ref{Figure1}).\\
For flickering experiments, movies are recorded across the equatorial plane. Details of the flickering analysis are reported in the next section. To improve the quality of the signal, imaging for morphological characterisation and flickering analysis is carried out on samples stained with Texas Red DHPE.\\
FRET spectroscopy measurements are carried out by performing a spectral emission scan ($\lambda$-scan) of the contact area between an adhering GUV and the surrounding free SBL. While exciting the donor (Cy3), the emission of donor and acceptor is reconstructed by scanning the acquisition window from 530 to 785 nm, with a 6.75 nm binning. Similarly to the case of z-stacks, the adhesion patch is identified in each image by filtering and thresholding. For FRET imaging, non-fluorescent lipids are used in order to prevent undesired energy transfer between the bilayer and the DNA tethers.\\

\subsubsection{Flickering analysis}

\begin{figure}[ht!]
\centering
  \includegraphics[width=7cm]{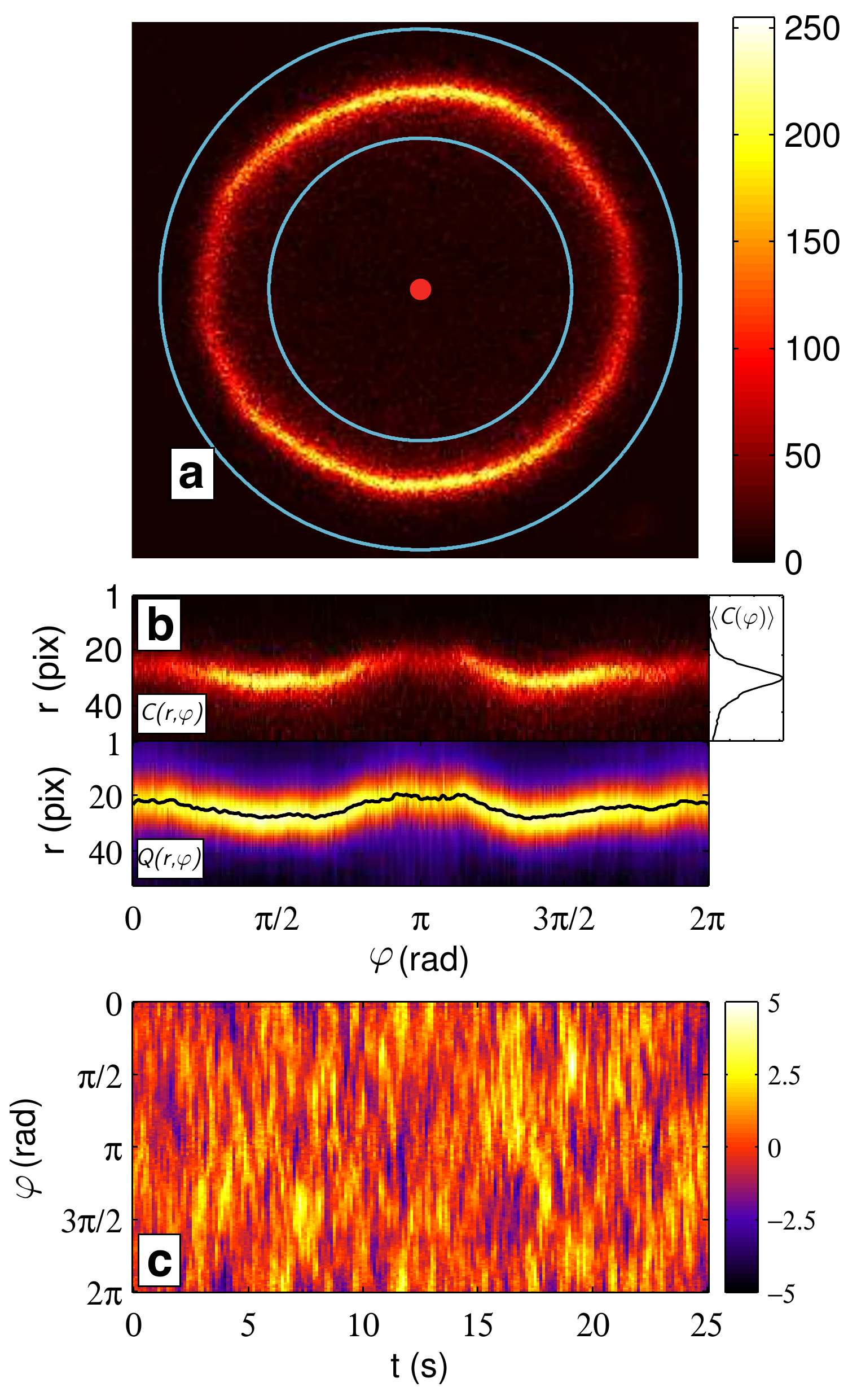}
  \caption{Intermediate steps of the contour detection algorithm. \textbf{a}, The position of the centra and the mean radius $R$ of the vesicle define an annular region in the frame;
\textbf{b}, the annular region $C(r,\varphi)$ extrapolated from the frame is used to compute an averaged template $\langle C(r) \rangle$; a 2D correlation $Q(r,\phi)$ of the two is computed and fitted to find the membrane position with sub-pixel precision, as described in the text (steps 1-6). The resulting contour is reported as a black line; \textbf{c}, colormap representation of the membrane fluctuations relative to the average radius. The colorbar reports the relative fluctuation as percentage in the range $-5\%$, $+5\%$ }
  \label{FigureFlickering}
\end{figure}

Typically, in case of phase-contrast imaging,  the contour of fluctuating GUVs is reconstructed by finding the inflection point in the radial intensity profile, as in reference.\cite{Pecreaux_EPJE_2004} In case of fluorescence imaging, the maximum of the intensity profile is commonly used to mark the membrane position. 
Here, we designed a Matlab algorithm to reconstruct the contour of GUVs from their equatorial cross section with sub-pixel precision, independently on the imaging mode employed.
The algorithm proceeds following these steps:
\begin{enumerate}
		\item The position of the centre, and a rough-guess value for the radius $R$ of the GUV are manually selected by the user on the first frame of the video (see Fig.~\ref{FigureFlickering}a).
		\item The image of the radial profile $C(r,\varphi)$ is selected within an annular region of user-defined width that contains the membrane. The annular region is then mapped onto a rectangular stripe using a cubic interpolation (see Fig.~\ref{FigureFlickering}b, top).
		\item A template $\langle C(r) \rangle$ of the radial intensity profile is calculated from the annular region, either by manual selection, or by automatically averaging along $\varphi$ (see Fig.~\ref{FigureFlickering}b, top-right).
		\item The algorithm computes the 2D correlation function $Q(r,\varphi)$ of $C(r,\varphi)$ with the template $\langle C(r) \rangle$. For each value of the polar angle $\varphi$, $Q(r,\varphi)$ is maximum at the radial coordinate $r$ where the template best correlates with the membrane profile (see Fig.~\ref{FigureFlickering}b, bottom).
		\item A rough measurement of the contour position $r(\varphi)$ is obtained by fitting the peak of the correlation image with a parabola around its maximum, for every value of $\varphi$. The algorithm cycles three times from point 2 to 5 , each time refining the estimate of the vesicle's centre and mean radius using the obtained contour.
		\item The rough contour estimate is used as the initial point of a global minimisation of the energy-like functional
		\begin{equation}
		\epsilon (r, \varphi) =  \sum_{\varphi} \left[ \left[ 1 - Q \left( r(\varphi),\varphi \right) \right] + \alpha \frac{\partial^2 r(\varphi) }{\partial \varphi^2} \right]
		\end{equation}
		here, the membrane position corresponds to the minimum of $1-Q$, corrected by a term that gives low weight to high curvature features, commonly due to random noise in the image. This retrieves the measurement of the contour position $r(\varphi)$. The minimisation makes use of the Nelder-Mead algorithm\footnote{Matlab's \textit{fminsearch} function}. 
\end{enumerate}
The procedure is repeated for every frame in a video of the fluctuating membrane, each time using the center and radius values of the previous frame as the starting point of the algorithm. An example of the temporal evolution of the membrane radial profile $r(\varphi)$ is shown in Fig.~\ref{FigureFlickering}c.
The spectrum of the thermal fluctuations of the contour is calculated using Matlab Fast Fourier Transform algorithm, using the formula
$$ \langle |h^{2} (q_x) | \rangle = \frac{ \langle \left| FFT \left ( R(\varphi,t) \right ) \right|^{2} \rangle_{t}}{M^{2}}$$
where M is the number points used to map the contour, and then averaged over the ensemble of frames.

\subsubsection{Control experiments}
Control experiments are performed to measure temperature-dependent membrane tension of non-adhering GUVs. Plain, non-functionalised, GUVs are imaged while laying on a glass substrate passivated with bovine serum albumin (BSA, Sigma Aldrich). Part of the images is carried out in bright-field microscopy, using a Nikon Eclipse Ti-E inverted microscope, a Nikon PLAN APO 40$\times$ 0.95 N.A. dry objective and a IIDC Point Grey Research Grasshopper-3 camera. For control experiments GUVs are diluted  in a 1:9 ratio in iso-osmolar glucose solution to enable sedimentation.


\section{Theoretical model}
We present a quantitative model describing the DNA-mediated adhesion of GUVs on flat supported bilayers. The model is adapted from reference,\cite{Parolini_NatComms_2015} where we treated the case of two identical adhering GUVs. Let us consider the interaction between an infinite SBL and a GUV adhering to it
\begin{equation}\label{U1}
U=U_\mathrm{membrane}+U_\mathrm{DNA}+U_0,
\end{equation}
where $U_\mathrm{membrane}$ accounts for the mechanical deformation of the GUV, $U_\mathrm{DNA}$ encodes for DNA-mediated adhesion, and $U_0$ is the reference energy, calculated for isolated GUV and SBL. 

\subsection{Membrane deformation}
In Eq.~\ref{U1}, $U_\mathrm{membrane}$ summarises three contributions: stretching energy, bending energy, and the entropic cost of suppressing thermal fluctuations of the contact area between {}the GUV and the SBL.\cite{Ramachandran_Langmuir_2010} In the limit of \emph{strong adhesion} the stretching energy dominates over the other two contributions, which we can neglect.\cite{Ramachandran_Langmuir_2010} As discussed later, for the case of DNA-mediated interactions, this condition is generally verified at low enough temperature. We can rewrite
\begin{equation}\label{U2}
U_\mathrm{membrane}(\theta)=U_\mathrm{stretching}(\theta)=K_\mathrm{a}\frac{\left[A(\theta)-\tilde{A}\right]^2}{2\tilde{A}},
\end{equation}
where $K_\mathrm{a}$ is the stretching modulus of the membrane, $A(\theta)$ is the overall (stretched) area of the GUV, and $\tilde{A}$ is the reference, un-stretched, area. In the limit of strong adhesion the GUV will take the shape of a truncated sphere with contact angle $\theta$ (Fig.~\ref{Figure1}), which we take as the independent variable of our model.\cite{Ramachandran_Langmuir_2010} Within the assumption of constant inner volume $V=4\pi R_0^3/3$, where $R_0$ is a reference radius, the total area $A$ of the GUV and the adhesion patch area $A_\mathrm{p}$ can be expressed as a function of the contact angle $\theta$(see Appendix, section \ref{App_geometry}). The un-stretched vesicle area $\tilde{A}$ exhibits a strong temperature-dependence
\begin{equation}\label{eqn:unstretched}
\tilde{A}=A_0\left[1+\alpha(T-T_0)\right],
\end{equation} 
where $\alpha$ is the area thermal expansion coefficient\cite{Evans_JPC_1987} and $T_0$ is the \emph{neutral temperature} of the GUVs, at which its reduced volume equals unity and $A=4\pi R_0^2$ (see Appendix, section \ref{App_geometry}).

\subsection{DNA-mediated adhesion}
We now focus on the DNA-mediated contribution to the interaction energy in Eq.~\ref{U1}: $U_\mathrm{DNA}$.\\ 
Given that the persistence length of dsDNA is $\sim50$~nm~$\gg$~$L=9.8$~nm,\cite{Smith_Science_1996} we can model the dsDNA spacers as rigid rods that, thanks to the fluidity of the membrane, can freely diffuse on the surface of the bilayers.\cite{Parolini_NatComms_2015,Angioletti-Uberti_PRL_2014} Free pivoting motion is guaranteed by the flexibility of the joint between the cholesterol anchor and the dsDNA spacer (Fig.~\ref{Figure1}). As demonstrated in ref.\cite{Parolini_NatComms_2015}, we can regard the sticky ends as point-like reactive sites, neglecting their physical dimensions. Moreover we can safely assume that the distance between the adhering membranes within the contact area is equal to $L$.\cite{Parolini_NatComms_2015} Finally, we neglect excluded volume interactions between unbound DNA tethers.\cite{Bracha_PNAS_2013,Parolini_NatComms_2015} The last two assumptions guarantee a uniform distribution of unbound DNA tethers and loops over the GUV and SBL surfaces.\\
The free energy change associated to the formation of a single bridge (b) or a loop ($\ell$) within the GUV is
\begin{equation}\label{eqn:singleenergy}
\Delta G_{\mathrm{b}/\ell}=\Delta G^0 - T \Delta S^\mathrm{conf}_{\mathrm{b}/\ell},
\end{equation}
where $\Delta G^0=\Delta H^0 - T \Delta S^0$ is the hybridisation free energy of untethered sticky ends, which can be calculated from the nearest-neighbour thermodynamic model,\cite{SantaLucia_PNAS_1998} eventually corrected to account for neighbouring non-hybridised bases.\cite{Di-Michele_JACS_2014} In Eq.~\ref{eqn:singleenergy}, the term  $-T\Delta S^\mathrm{conf}_{\mathrm{b}/\ell}$ accounts for the confinement entropic loss taking place when tethered sticky ends hybridise,\cite{Angioletti-Uberti_NMat_2012,Mognetti_SoftMatter_2012,Varilly_JChemPhys_2012} and can be estimated as (see Appendix, section \ref{App_FreeEnergy})
\begin{equation}\label{DeltaG}
\Delta G(\theta) =\Delta G^0 - k_\mathrm{B}T\log\left[\frac{1}{\rho_0 L A(\theta)}\right],
\end{equation}
where $\rho_0=1$M is a reference concentration, and we highlighted the coupling with the geometry of the GUV via the $\theta$-dependence. Note that, contrary to the case of two adhering GUVs,\cite{Parolini_NatComms_2015} here $\Delta S^\mathrm{conf}_\mathrm{b}=\Delta S^\mathrm{conf}_\ell$. The local roughness of the membranes could also influence $\Delta G$\cite{Hu_PNAS_2013} -- this effects will be studied elsewhere.\\
We indicate with $N$ the total number of $a$ and $a'$ tethers on the GUVs, and model the SBL as an infinite reservoir of tethers. A combinatorial calculation detailed in the Appendix (sections \ref{App_SBL} and \ref{App_Comb}) allows the derivation of the overall hybridisation energy for the system of linkers
\begin{equation}\label{UHyb}
U_\mathrm{hyb}=N k_\mathrm{B} T \left[x_\ell + 2 \log \left(1-x_\ell-x_\mathrm{b}\right)-2\frac{\bar{N}_\mathrm{f}}{N}\right],
\end{equation}
In Eq.~\ref{UHyb}, $x_\mathrm{b}$ and $x_\ell$ are the fraction of tethers involved in bridges/loops, given by
\begin{equation}\label{BridgesImpl}
\frac{x_\mathrm{b}}{\left(1-x_\ell-x_\mathrm{b}\right)}=q_\mathrm{b},
\end{equation}
\begin{equation}\label{LoopsImpl}
\frac{x_\ell}{\left(1-x_\ell-x_\mathrm{b}\right)^2}=q_\ell
\end{equation}
where
\begin{eqnarray}
q_\ell&=&\exp\left[-\beta \Delta G^*\right], \label{QLoops}\\
q_\mathrm{b}&=&\frac{\bar{N}_\mathrm{f}}{N}\exp\left[-\beta \Delta G^*\right].\label{QBridges}
\end{eqnarray}
In Eqs.~\ref{QLoops} and \ref{QBridges} the hybridisation free energy is re-defined as $\Delta G^* = \Delta G - k_\mathrm{B}T \log N$. The combinatorial contribution $-k_\mathrm{B}T \log N$, typically estimated in $\sim-10 k_\mathrm{B}T$, has a stabilising effect.\\
The quantity $\bar{N}_\mathrm{f}$ appearing in Eq.~\ref{UHyb} indicates the average number of unbound DNA tethers anchored to the SBL available within the adhesion patch. The concentration $c_\mathrm{f}$ of unbound tetheres of each type ($a$ or $a'$), such that $\bar{N}_\mathrm{f}(\theta)=c_\mathrm{f}A_\mathrm{p}(\theta)$, is given by
\begin{equation}\label{eqn:freeSBL}
c_0-c_\mathrm{f}=c_\mathrm{f}^2\frac{\exp[-\beta \Delta G_0]}{\rho_0 L},
\end{equation}
where $c_0=\rho_\mathrm{DNA}/2$ is the total concentration of $a$ and $a'$ tethers. The full derivation of Eq.~\ref{eqn:freeSBL} is provided in the Appendix (section \ref{App_Comb}). Note that the concentration of loops on the SBL is $c_\ell=c_0-c_\mathrm{f}$.\\
Equation \ref{UHyb} generalises the expression found in ref.\cite{Parolini_NatComms_2015} to the case in which vesicles are in contact with an infinite reservoir of thethers.
The overall DNA contribution to the interaction energy in Eq.~\ref{U1} is
\begin{equation}\label{UDNA}
U_\mathrm{DNA}=U_\mathrm{hyb}-2N k_\mathrm{B} T \log\left(\frac{A}{A_0}\right),
\end{equation}
where the term on the right-hand side accounts for the change in overall confinement entropy following the area-change of the GUV.\cite{Parolini_NatComms_2015}\\

\subsection{Overall interaction energy.~~}
By combining Eqs.~\ref{U2}, \ref{UHyb}, and \ref{UDNA} into Eqs.~ \ref{U1}, we obtain an analytical expression for the interaction energy of the GUV+SBL system 
as a function of the only independent variable $\theta$
\begin{multline}
U(\theta)-U_0=K_\mathrm{a}\frac{\left[A(\theta)-\tilde{A}\right]^2}{2\tilde{A}}+ N k_\mathrm{B} T \Bigl[x_\ell(\theta) + \\
 2 \log \left[1-x_\ell(\theta)-x_\mathrm{b}(\theta)\right]-2\frac{\bar{N}_\mathrm{f}(\theta)}{N} - 2 \log\left(\frac{A(\theta)}{A_0}\right) \Bigr].
\end{multline}

Note that $U$ depends on $\theta$ though the adhesion area $A_\mathrm{p}$ (see Eq.~\ref{Nbar}) and the total area $A$ of the GUV. On the other hand, the reference energy $U_0$ is $\theta$-independent (see derivation in the Appendix, section \ref{App_Ref}). The interaction energy can be minimised to calculate all the morphological observables (e.g. the contact angle $\theta$, area of the adhesion patch $A_\mathrm{p}$, area of the spherical section of the GUV $A$), as well as the fraction of formed bridges and loops ($x_{\mathrm{b}/\ell}$).

\begin{figure}[ht!]
\centering
  \includegraphics[width=8cm]{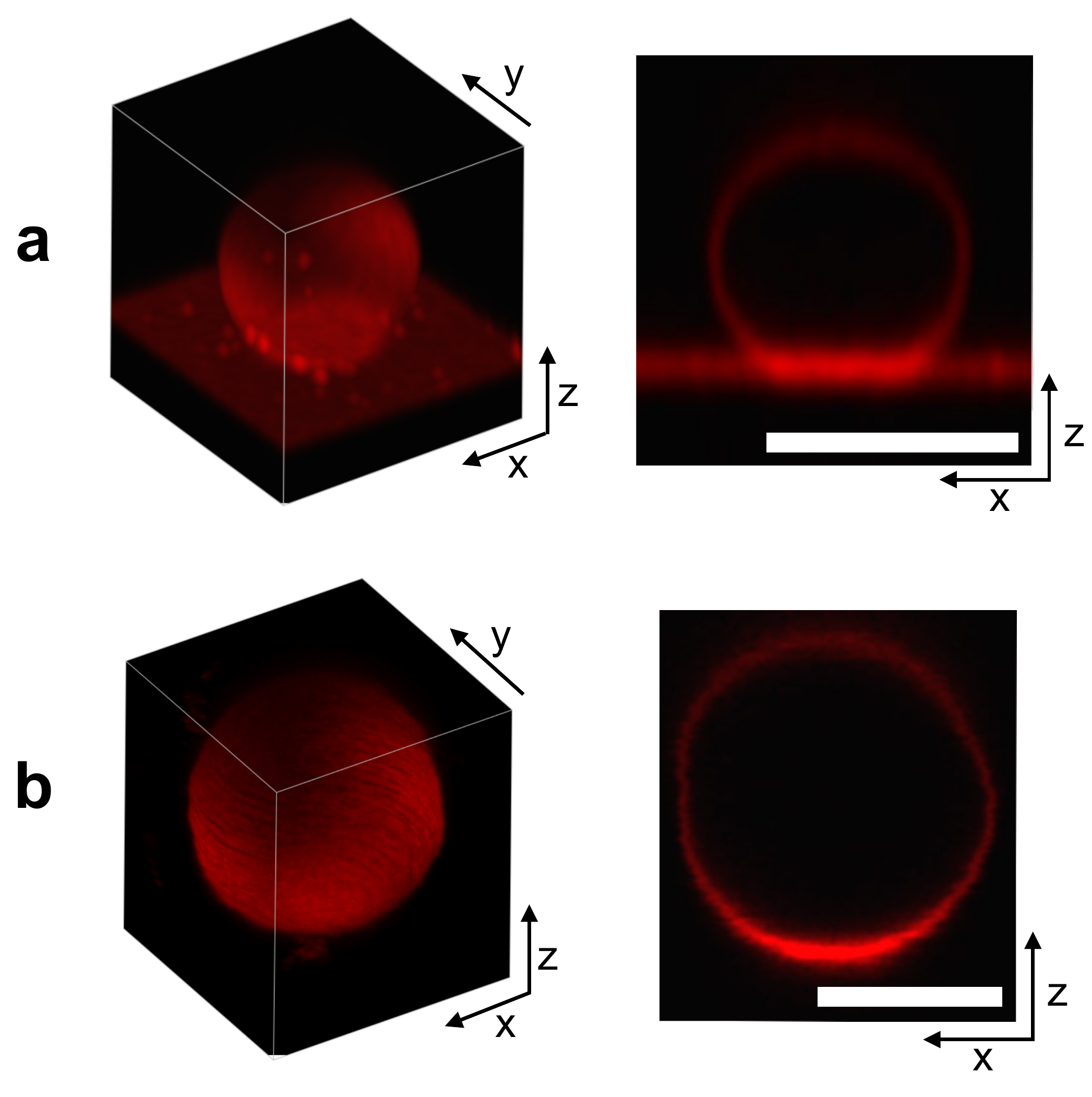}
  \caption{From left to right: 3D reconstruction from confocal z-stack, and confocal cross section of an adhering GUV (\textbf{a}) and a non-adhering GUV on a glass substrate (\textbf{b}). For 3D rendering, confocal images have been acquired with a Leica HCX PL APO CS 63$\times$ 1.4 NA oil immersion objective for better resolution, and deconvolved using the experimental point-spread function. Scale bars: $10$ $\mu$m.}
  \label{FigureStacks}
\end{figure}

\subsection{Model parameters and error propagation}
The model features seven input parameters: the thermal expansion coefficient $\alpha$, the stretching modulus $K_\mathrm{a}$, the length of the dsDNA tether $L$, the hybridisation enthalpy $\Delta H^0$ and entropy $\Delta S^0$  of the sticky ends, the DNA coating density $\rho_\mathrm{DNA}$ (used to calculate the overall number of strands per GUV -- $2N$), the neutral temperature $T_0$ and radius $R_0$.\\ The stretching modulus is estimated from literature data as $K_\mathrm{a} = 240\pm90$ mN~m$^{-1}$, with the error bar covering the entire range of reported values.\cite{Rawicz_BJ_2000,Rawicz_BJ_2008,Fa_BBA_2007,Pan_BJ_2008} The thermal expansion coefficient has been experimentally estimated as $\alpha=1.3\pm0.7$ K$^{-1}$.\cite{Parolini_NatComms_2015} The hybridisation enthalpy and entropy are estimated according to the nearest neighbours  thermodynamic rules\cite{SantaLucia_PNAS_1998} as $\Delta H^0=-54\pm5$ kcal~mol$^{-1}$ and $\Delta S^0=-154\pm13$ cal~mol$^{-1}$~K$^{-1}$. It is not clear whether the stabilising effect of the non-hybridised dangling bases neighbouring the sticky ends, in the present case the adenine bases present at both sides of the sequences (see Fig.~\ref{Figure1}), should be taken into account. It is indeed possible that their attractive contribution is compensated  or overwhelmed by the repulsive effect of the long inert DNA connected to the sticky ends.\cite{Di-Michele_JACS_2014} The errorbars in $\Delta H^0$ and $\Delta S^0$ are included to cover both these scenarios. The length $L=9.8$~nm of the dsDNA spacers can be precisely estimated from the contour length of dsDNA ($0.388$ nm per base-pair\cite{Smith_Science_1996}). The DNA coating density is estimated for different samples as explained in the experimental section. The neutral temperature $T_0$ changes widely from vesicle to vesicle due to the polydispersity of electroformed samples. We generally use $T_0$ as a fitting parameter. For a known $T_0$, the neutral radius $R_0$ is experimentally determined as $R_0=R|_{T=T_0}$. If $T_0$ falls outside the experimentally accessible temperature range, we extrapolate 
\begin{equation}
R_0\approx \sqrt{\frac{A(T_1)}{\pi \left[1+\alpha\left(T_1-T_0\right)\right]}},
\end{equation}
where $T_1$ is the minimum experimentally accessible temperature.\\
Errors in the input parameters are numerically propagated to the theoretical predictions.\cite{Parolini_NatComms_2015} Briefly, we sample the results of the model using random values of the input parameters $X\pm \Delta  X$ extracted from a Gaussian distribution with mean equal to $X$ and standard deviation equal to $\Delta X$. From a sample of size $\ge10000$, we estimate the theoretical prediction of each observable as the median of the sampled distribution. Errobars cover the interval between the $16^\mathrm{th}$ and the $84^\mathrm{th}$ percentile. 

\begin{table}[ht!]
\small
\caption{\textbf{Input parameters of the model}}  
\label{Table}
  \begin{tabular*}{0.5\textwidth}{@{\extracolsep{\fill}}ccc}
    \hline
& $\alpha = 1.3\pm0.7$ K$^{-1}$ & \\
& $K_\mathrm{a} = 240\pm90$ mN~m$^{-1}$ &\\
& $L=9.8$ nm &\\
& $\Delta H^0=-54\pm5$ kcal~mol$^{-1}$ &\\
& $\Delta S^0=-154\pm13$ cal~mol$^{-1}$~K$^{-1}$& \\
    \hline
  \end{tabular*}
\end{table}


\begin{figure}[ht!]
\centering
  \includegraphics[width=8.8cm]{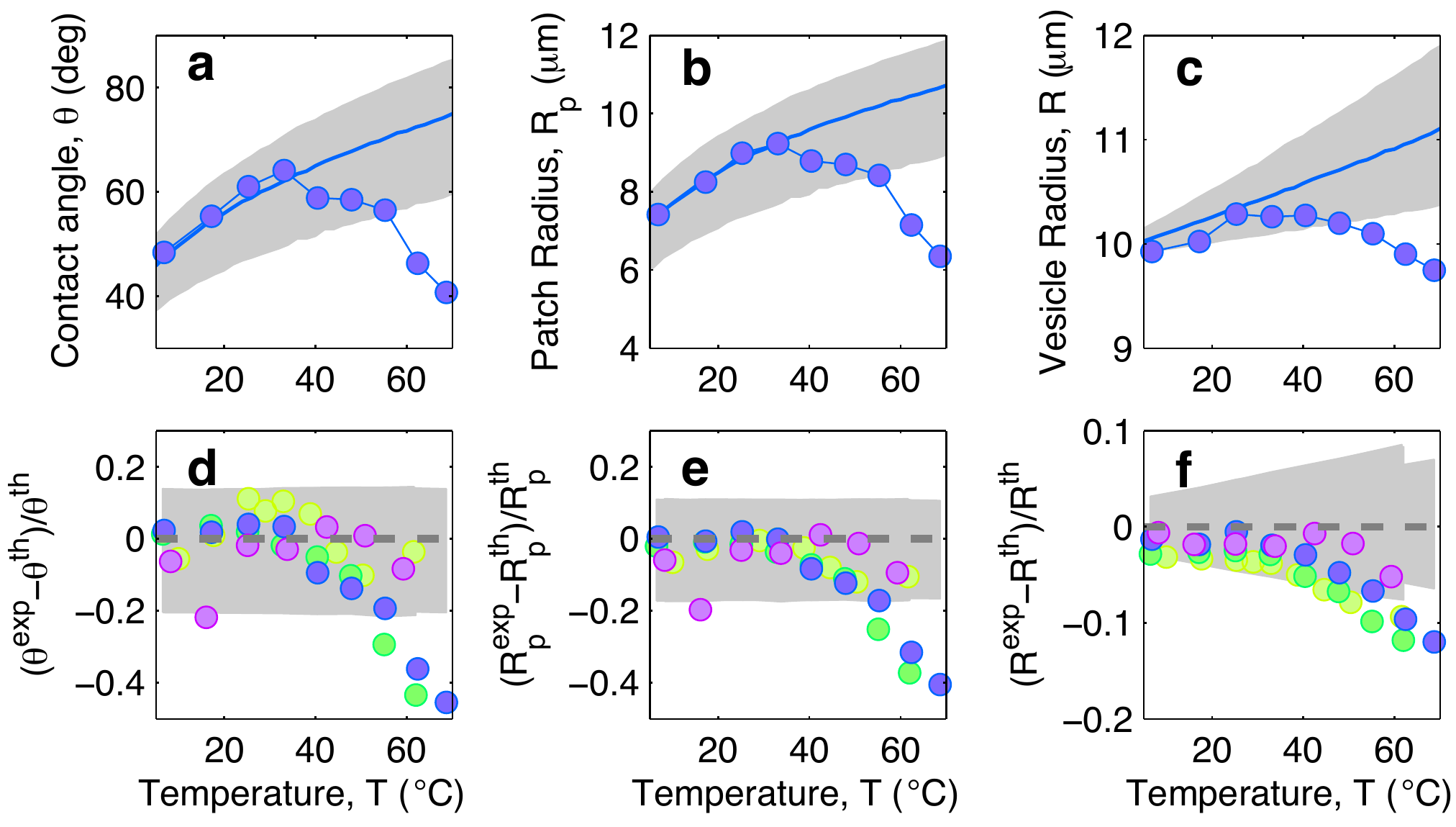}
  \caption{Experimental and theoretical temperature-dependent vesicle adhesion for samples with intermediate DNA coating density $\rho_\mathrm{DNA}^\mathrm{med}=390\pm90$ $\mu$m$^{-2}$ (see experimental section). In the top row we demonstrate the temperature-dependence of the contact angle $\theta$ (\textbf{a}), adhesion-patch radius $R_\mathrm{p}$ (\textbf{b}), and vesicle radius $R$ (\textbf{c}) for a typical adhering GUV. Points indicate experimental data, solid lines mark theoretical predictions, with errorbars visualised as grey-shaded regions. Fitting parameter $T_0=-40^\circ$C. In the bottom row we summarise the results for 5 vesicles. Points represent the relative deviation of experimental data from theoretical predictions $(X^\mathrm{exp}-X^\mathrm{th})/X^\mathrm{th}$, with $X=\theta$ (\textbf{d}), $R_\mathrm{p}$ (\textbf{e}), and $R$ (\textbf{f}). Grey-shaded regions mark the uncertainty interval of the theory. Model parameters are reported in Table \ref{Table}.}
  \label{FigureShapeHigh}
\end{figure}

\section{Results and discussion}
\subsection{Qualitative observations}
In Fig.~\ref{FigureStacks} we can visually compare confocal images of a DNA-functionalised GUV adhering to a SBL (a) and a non-adhering GUV on a passivated glass surface (b). Both GUVs and SBL are stained with fluorescent lipids.\\
It is clear from both the 3D reconstruction and the vertical cross section that the adhering GUV takes the shape of a truncated sphere, with a flat and circular contact region. This evidence confirms the assumption that, at low enough temperature, DNA-mediated adhesion is strong enough to guarantee the dominance of stretching over bending and other contributions to the deformation energy. The fluorescence intensity measured within the adhesion patches is almost exactly equal to twice the value measured on the SBL outside the adhesion region (1.95 times for the vesicle shown in Fig.~\ref{FigureStacks}a). This evidence confirms the presence of two lipid bilayers in close contact within the adhesion area and excludes the possibility of DNA-mediated fusion of the two membranes.\cite{Stengel_JCPB_2008}\\ The non-adhering GUV displayed in Fig.~\ref{FigureStacks}b does not exhibit a flat adhesion patch, as clear from the vertical cross-section. Note that for both the adhering and the non-adhering GUVs, the bottom part of the stacks appears brighter due to the z-dependent response of the instrument.\\

\begin{figure}[ht!]
\centering
  \includegraphics[width=8.8cm]{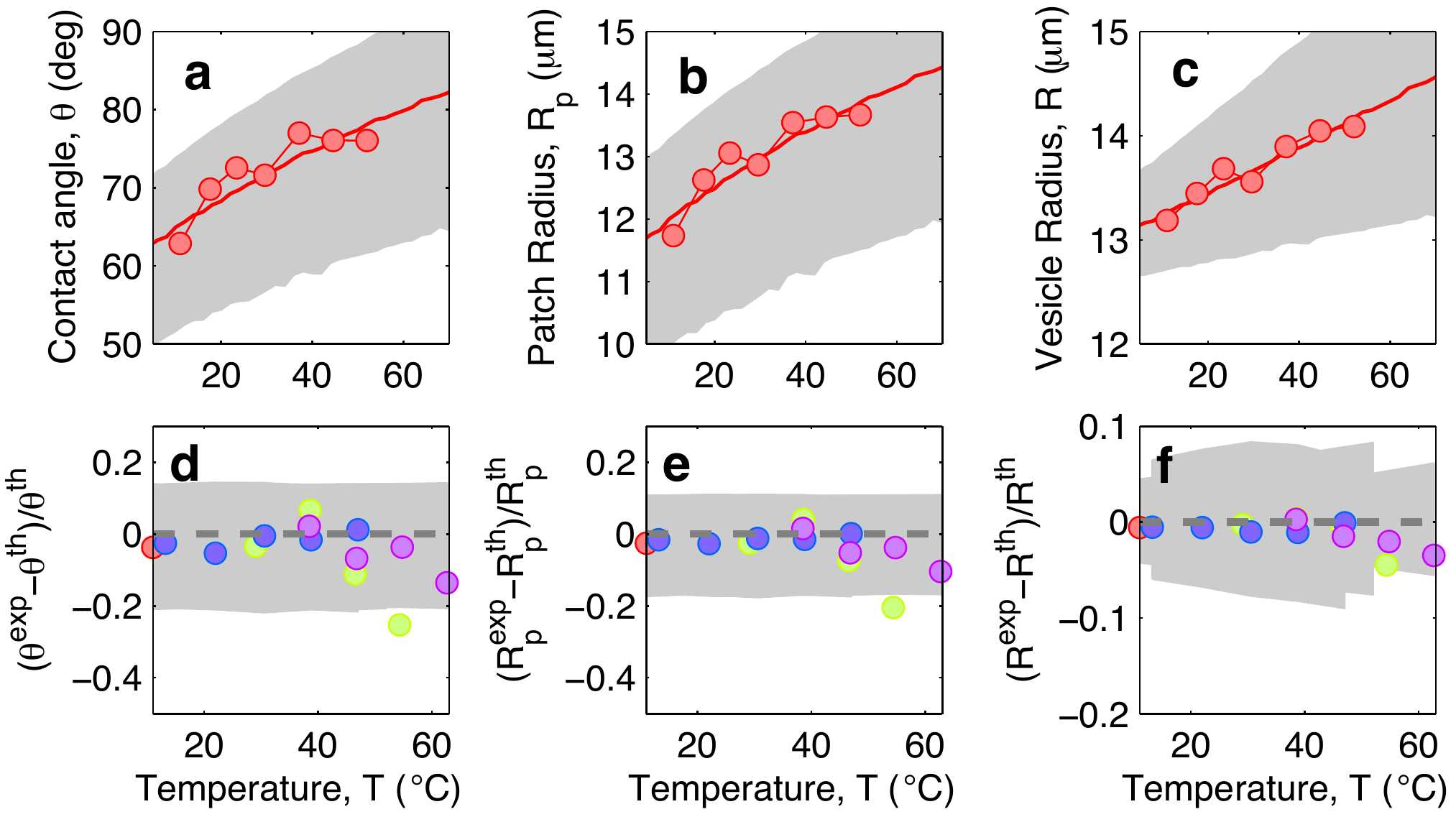}
  \caption{Experimental and theoretical temperature-dependent vesicle adhesion for samples with low DNA coating density $\rho_\mathrm{DNA}^\mathrm{low}=39\pm9$ $\mu$m$^{-2}$ (see experimental section). In the top row we demonstrate the temperature-dependence of the contact angle $\theta$ (\textbf{a}), adhesion-patch radius $R_\mathrm{p}$ (\textbf{b}), and vesicle radius $R$ (\textbf{c}) for a typical adhering GUV. Points indicate experimental data, solid lines mark theoretical predictions, with errorbars visualised as grey-shaded regions. Fitting parameter $T_0=-20^\circ$C. In the bottom row we summarise the results for 5 vesicles. Points represent the relative deviation of experimental data from theoretical predictions $(X^\mathrm{exp}-X^\mathrm{th})/X^\mathrm{th}$ for $X=\theta$ (\textbf{d}), $R_\mathrm{p}$ (\textbf{e}), and $R$ (\textbf{f}). Grey-shaded regions mark the uncertainty interval of the theory.  Model parameters are reported in Table \ref{Table}.}
  \label{FigureShapeLow}
\end{figure}

\subsection{Temperature-dependence of the geometrical observables}
In this section we discuss the temperature-dependence of the morphology of adhering GUVs. In Fig.~\ref{FigureShapeHigh}a-c we show the experimentally determined contact angle $\theta$, adhesion-patch radius $R_\mathrm{p}$, and vesicle radius $R$ for a typical adhering GUV is a sample with DNA concentration equal to $\rho_\mathrm{DNA}^\mathrm{med}=390\pm90$ $\mu$m$^{-2}$. The contact angle displays a non-monotonic behaviour as a function of temperature, with a positive slope at low $T$, followed by a sudden decay for $T$ higher than $\sim30^\circ$C. The adhesion radius follows the same trend, as does the vesicle radius, which however displays much smaller relative variations. The solid lines in Fig.~\ref{FigureShapeHigh}a-c represent theoretical predictions calculated using the input parameters in Table.~\ref{Table}, and using the neutral temperature $T_0$ as a fitting parameter. Grey-shaded regions indicate propagated uncertainty in the theoretical predictions. The agreement between theory and experiments is quantitative at low temperatures. At high $T$, the theory fails to predict the drop in contact angle observed in experiments. This behaviour is expected since our theoretical description is valid in the limit of strong adhesion, where the attractive forces are sufficient to suppress bending contributions in the interaction energy. At high temperature the DNA, which in the present experiment features relatively short sticky ends, starts to melt, causing the loosening of the adhesive forces and a change in the GUV shape, detected as a shrinkage of the adhesion area. The temperature-dependence of the fraction of DNA bonds is quantified and discussed in the following sections.\\ In Fig.~\ref{FigureShapeHigh}d-f we show the relative deviations of the experimentally-determined morphological observables from the theoretical predictions, defined as $(X^\mathrm{exp}-X^\mathrm{th})/X^\mathrm{th}$, for $X=\theta$, $A_\mathrm{p}$, and $A$. The data, collected from 5 vesicles are consistent: the experimental data fall within theoretical errorbars at low $T$, deviating at higher temperature due to the failure of the strong adhesion assumption.\\ 
In Fig.~\ref{FigureShapeLow} we show experimental, and theoretically predicted morphological observables for the case of low DNA concentration, $\rho_\mathrm{DNA}^\mathrm{low}=39\pm9$ $\mu$m$^{-2}$. Similarly to the case of higher DNA concentration, experimental data are backed by theoretical predictions. In this case, however, the high-temperature deviation of the experiments from the theoretical predictions appears to be less evident, and shifted towards higher temperatures. This indicates the presence of an (however small) adhesive force hindering the partial detachment of the GUVs. We ascribe this behaviour to the effect of non-specific membrane-membrane adhesion, e.g. dispersion attraction, that for higher DNA coverage is suppressed by steric repulsion.\\

\begin{figure*}[ht!]
\centering
  \includegraphics[width=14cm]{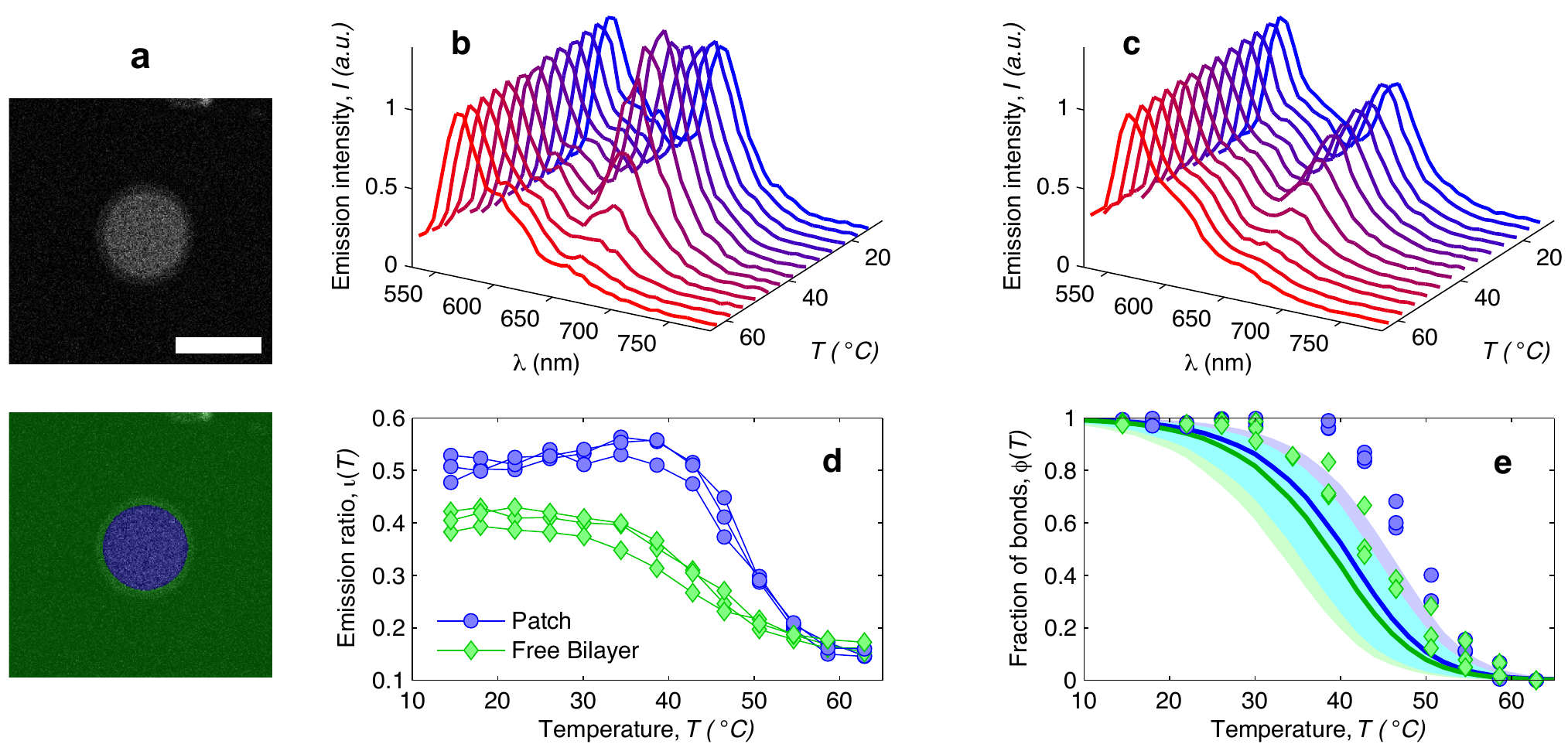}
  \caption{In-situ FRET spectroscopy characterisation of the temperature-dependence of the fraction of DNA bonds. \textbf{a}, Confocal image of the adhesion patch of a GUV. At the bottom we show the same image segmented with our software to separate the adhesion patch (blue) from the free supported bilayer (green). Scale bar: 15~$\mu$m. \textbf{b}, Fluorescence emission spectra measured within the adhesion patch for a typical vesicle (blue area in panel \textbf{a}). Colours from blue to red mark low to high temperatures in the interval $14.5\le T \le 62.9^\circ$C. The curves are normalised to the emission peak of Cy3 ($568$nm). In panel \textbf{c} we show the emission spectra measured on the free SBL (green area in panel \textbf{a}). \textbf{d}, Relative intensity of the Cy5 emission peak ($665$nm) measured within (blue circles) and outside (green lozenges) the adhesion patch. The curves are relative to 3 different vesicles. The amplitude of the Cy5 and Cy3 emission peaks is determined through a Gaussian fit of the 5 data points closest to the maximum. \textbf{e}, Experimental (symbols) and theoretically predicted (solid lines) fraction of formed DNA bonds within (blue circles) and outside (green lozenges) the adhesion patch. Experimental data are extracted from the curves in panel \textbf{d} as described in the text. Theoretical curves are calculated using the parameters in Table~\ref{Table}, DNA-coating density $\rho^\mathrm{high}_\mathrm{DNA}=1200\pm300$ $\mu$m$^{-2}$, $T_0=-20$, and $R_\mathrm{0}=10$ $\mu$m. Note that the value of $T_0$ does not significantly affect these quantities. Blue and Green shaded regions indicate propagated uncertainties of the blue and green solid lines. Cyan shaded region marks their overlap.}
  \label{FigureFRET}
\end{figure*}

\subsection{DNA-melting}
We investigate the temperature-dependence of the fraction of formed DNA bonds via in-situ FRET measurements. Cyanine fluorophores, Cy3 (donor) and Cy5 (acceptor), are connected to the $3'$ termini of $a$ and $a'$ sticky ends, as sketched in Fig.~\ref{Figure1}. FRET efficiency is described by 
\begin{equation}
E=\frac{1}{1+\left(\frac{d}{R_0}\right)^6},
\end{equation}
where $d$ is the distance between the fluorohpores and  $R_0$ is the Forster radius, equal to $5.4$~nm for the case of Cy3-Cy5.\cite{Yuan_NAR_2007} When sticky ends are bound to form a loop or a bridge, the distance between Cy3 and Cy5 is approximately equal to the length of the hybridised sticky ends -- 2.7~nm, therefore we can assume a very high energy transfer efficiency for bound linkers. The average distance between unbound linkers is sufficiently high to guarantee a comparatively very low transfer efficiency between unpaired tethers. Note that for FRET experiments the lipid membrane is not stained with Texas Red to avoid spurious signal (energy transfer between Texas Red and Cy5). In Fig.~\ref{FigureFRET}a we show the confocal image of an adhesion patch (top), segmented to separate the actual adhesion area from the surrounding free SBL. This enables an efficient characterisation of FRET efficiency in-situ. The emission spectra, collected within the patch and on the SBL while exciting the donor at 514~nm, are shown in Fig.~\ref{FigureFRET}c and d respectively. Colours from blue to red indicate low to high temperature. Spectra are normalised to the emission peak of Cy3, correctly found at $\sim568$~nm. Note that for this experiment we used high DNA concentration $\rho_\mathrm{DNA}^\mathrm{high}$ to strengthen the signal that otherwise would be too weak for a wavelength scan. As expected, the emission of the acceptor, peaked at $\sim665$~nm, visibly drops at high temperature. We quantify this effect in Fig.~\ref{FigureFRET}d, where we plot the normalised acceptor emission intensity $\iota=I_\mathrm{Cy5}/\left(I_\mathrm{Cy5}+I_\mathrm{Cy3}\right)$,\cite{McCann_BJ_2010} where the $I_\mathrm{Cy5/Cy3}$ are the peak-intensities estimated through a local Gaussian fit. Although qualitatively similar, the $\iota$ curves measured within and outside the adhesion patch exhibit some differences. For the case of free SBL, the emission ratio remains constant ($\sim 0.4$) or slightly decreases upon heating, up to $\sim 40^\circ$C, then it gradually drops down to $\sim 0.15$. This decay is ascribed to the melting transition of DNA loops formed within the SBL. The FRET signal measured within the patch is higher at low temperatures ($\sim 0.6$). This effect is probably due to the higher DNA density found within the patch at low temperature, which increases the probability of energy transfer between unpaired strands. Indeed  we find that the overall fluorescence intensity measured within the patch at $T<20^\circ$C is between 6 and 12 times higher than the intensity measured on the SBL. The FRET signal measured within the patch is also found to increase upon heating, before suddenly decreasing at $T\sim45^\circ$C. The increase in FRET efficiency cannot be explained by an increase in DNA density within the patch, since the local DNA density decreases as the adhesion area becomes larger upon heating. A possible explanation of this behaviour could be radiative cross-excitation between the two fluorophores, that becomes more efficient as the adhesion patch gets less crowded upon heating. At high temperatures the FRET signal measured within the patch plateaus at $\sim 0.15$, in line with what we measure on the SBL. This confirms that at high enough temperature, when no bonds are formed, the DNA concentration is uniform across all the surfaces.\\
The curves $\iota(T)$ can be used to semi-quantitatively estimate the temperature dependence of the overall fraction of DNA bonds. We fit the low temperature plateaus ($T<35^\circ$C) in Fig.~\ref{FigureFRET}d with linear baselines $B(T)$ and assume that $\iota(T)$ plateaus to a constant value for $T>57^\circ$C. The fraction of formed DNA bonds is thereby estimated as
\begin{equation}
\phi(T)=\frac{\iota(T)-\langle \iota_{T>57^\circ\mbox{C}}\rangle}{B(T)-\langle \iota_{T>57^\circ\mbox{C}}\rangle}.
\end{equation}
A better estimate of $\phi(T)$ could be obtained by measuring $\iota(T)$ up to higher temperatures, and fitting the high-temperature plateau with a second linear baseline. However, temperatures higher than $65^\circ$C cannot be safely probed due to the risk of destabilisation of the dsDNA spacers. The experimental $\phi(T)$ data in Fig.~\ref{FigureFRET}e indicate that the DNA melting transition is relatively broad, spanning more than $30^\circ$C.\cite{Meulen_JACS_2013} Moreover, the melting seems to occur at a higher temperature (by about $5^\circ$C) within the adhesion patch.\\
The experimentally estimated fraction of DNA bonds can be compared with theoretical predictions. On the free SBL, only loops can form, and therefore $\phi(T)$ should be compared to the fraction of loops $x_\ell^\mathrm{SBL}$, calculated accordingly to Eqs.~\ref{eqn:freeSBL}, \ref{fractionofloopsSBL}-\ref{qloopsSBL}. Within the adhesion patch we count contributions from bridges, loops formed on the GUV, and loops formed on the SBL. By assuming evenly distributed loops, the overall number of DNA bonds found within the patch is
\begin{equation}
N_\mathrm{bound}=N\left(\frac{A_\mathrm{p}}{A}x_\ell+x_\mathrm{b}\right)+c_0 x_\ell^\mathrm{SBL} A_\mathrm{p},
\end{equation}
where $x_{\mathrm{b}/\ell}$ are the fractions of loops and bridges on the GUV, given by Eqs.~\ref{Bridges} and \ref{Loops}. The overall number of DNA tethers, bound and unbound, found within the patch is
\begin{equation}
N_\mathrm{tot}=N\left[\frac{A_\mathrm{p}}{A}\left(1-x_\mathrm{b}\right)+x_\mathrm{b}\right]+c_0A_\mathrm{p},
\end{equation}
where we assume that also unbound DNA is evenly distributed across the surfaces.
The theoretically predicted fraction of bonds within the patch is thus $N_\mathrm{bound}/N_\mathrm{tot}$. In Fig.~\ref{FigureFRET}e we compare theoretical $\phi(T)$ with experimental estimates. Since the choice of the neutral temperature $T_0$ and radius $R_0$ do not noticeably affect the melting curves, theoretical predictions are calculated using a fixed $T_0=-20^\circ$C and $R_0=10$ $\mu$m, with no fitting parameters. Our model captures the width of the DNA transition as well as the difference in melting temperature between the patch and the free SBL. However the theory underestimates the average melting point by $5-10^\circ$C. This deviation is, at least partially, ascribable to the attractive effect of Cy3 and Cy5 molecules on the stability of DNA. For duplexes labeled with either of the molecules, the stabilisation has been quantified in a positive melting-temperature shift of $1.4-1.5^\circ$C.\cite{Moreira_BBRC_2005} The presence of both molecules is expected to cause a greater shift. Another explanation could be an underestimation of the DNA concentration $\rho_\mathrm{DNA}^\mathrm{high}$. As discussed below, this hypothesis is consistent with tension measurements. The impossibility of probing the high-temperature baselines could also play a role.\\

\begin{figure}[ht!]
\centering
  \includegraphics[width=8.8cm]{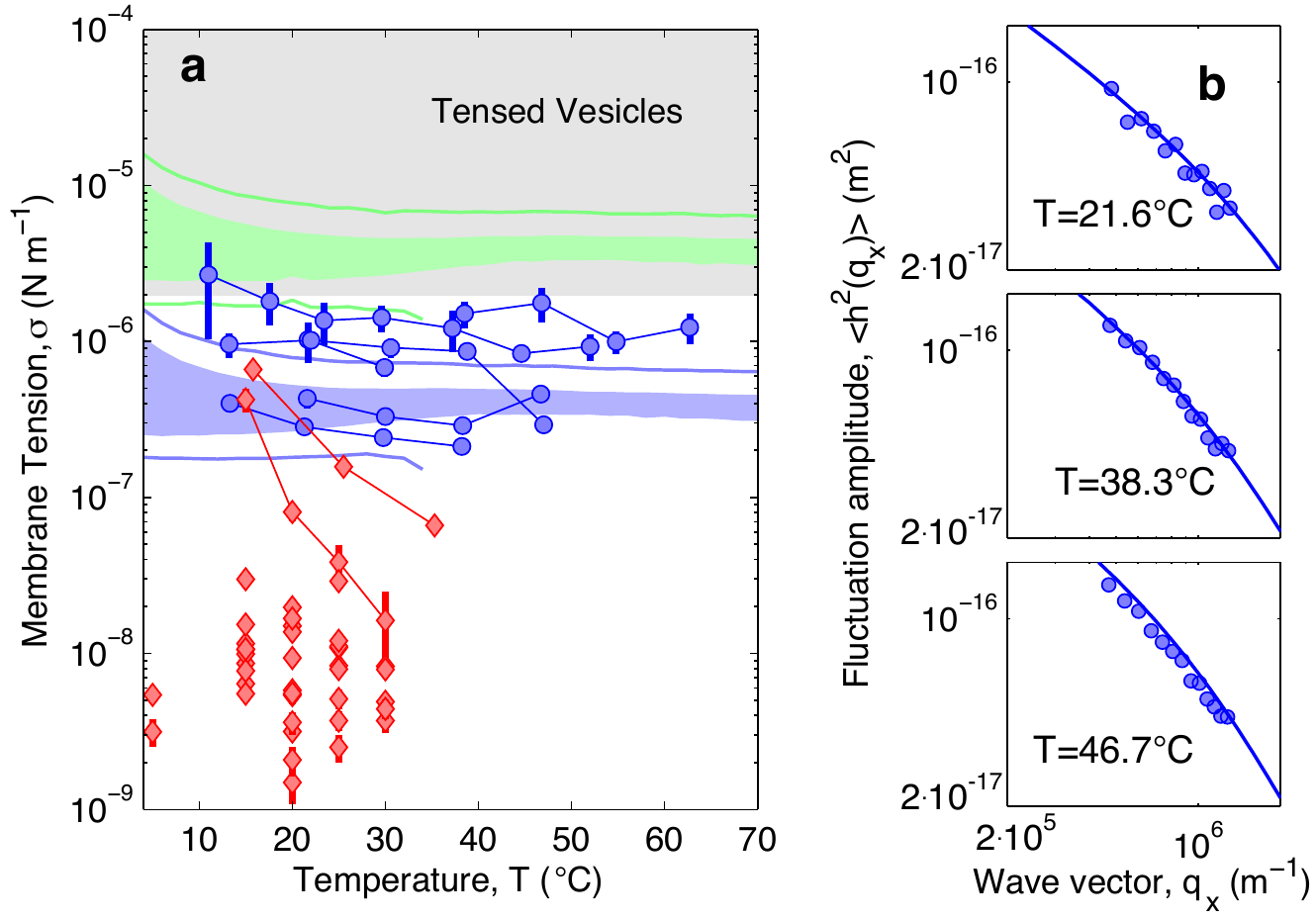}
  \caption{Experimental and theoretical temperature-dependent membrane tension. \textbf{a}, Membrane tension $\sigma$ measured from filckering experiments. Blue circles indicate adhering GUVs with low DNA coverage ($\rho_\mathrm{DNA}^\mathrm{low}=39\pm9$ $\mu$m$^{-2}$). Red lozengges indicate non-adhering GUVs on a BSA-coated glass surface. Grey-shaded regions mark regimes in which membrane tension is currently inaccessible to our technique. The blue-shaded band indicates the theoretical prediction for $\sigma$ calculated with the fitting parameter $-60\le T_0 \le 0^\circ$C and low DNA coverage. Blue lines mark the corresponding errorbars.  The green-shaded band indicate the theoretical prediction for $\sigma$ for higher DNA coverage ($\rho_\mathrm{DNA}^\mathrm{med}=390\pm90$  $\mu$m$^{-2}$). Green lines mark the corresponding errorbars. Model parameters are reported in Table \ref{Table}.\cite{Parolini_NatComms_2015} \textbf{b},~Fluctuation-amplitude spectra for adhering GUVs with low DNA coverage and various temperatures. Symbols indicate experimental data and solid lines indicate fits according to Eq.~\ref{eqn:fitspectrum}. From top to bottom the fitted values of the membrane tension are $\sigma=4.1\pm0.5$, $2.42\pm0.09$, $1.88\pm0.09\times10^{-7}$~Nm.}
  \label{FigureTension}
\end{figure}

\subsection{Membrane tension}
The temperature dependence of the membrane tension is measured by flickering analysis of the equatorial cross sections of GUVs. The tension is extracted by fitting the power spectra of the thermal fluctuations, determined as explained in the experimental section, with the function
\begin{equation}\label{eqn:fitspectrum}
\langle |h^2(q_x)| \rangle = \frac{k_\mathrm{B}T}{2 \mathcal{L} \sigma} \left[\frac{1}{q_x}-\frac{1}{\sqrt{q_x^2+\frac{\sigma}{\kappa}}}\right],
\end{equation}
where $\mathcal{L}$ is the contour length of the equatorial cross-section of the GUV, $\sigma$ is the membrane tension, $\kappa$ the bending modulus, and $q_x$ the wave vector evaluated along the contour.
Equation \ref{eqn:fitspectrum} is derived from the original work of Helfrich,\cite{Helfrich} describing the fluctuations of an infinite 2D membrane, and corrected to account for the fact that, by imaging an equatorial cross-sections, only modes propagating along the horizontal direction should  be considered.\cite{Pecreaux_EPJE_2004} Of the discrete set of wave vectors $q_x(n)=2 \pi n / \mathcal{L}$, modes with $n<6$ are excluded from the analysis. Mode $n=0$ and $n=1$ correspond to size changes and translations of the GUV. Modes with $n>2$ describe thermal fluctuations. However, Eq.~\ref{eqn:fitspectrum} is derived for a planar membrane, and should not be used to describe modes with $n<6$, which are influenced by the spherical geometry of the GUV.\cite{Pecreaux_EPJE_2004} For our analysis we fit the spectra for modes $6\le n \le 16$. At higher $q$ we approach the resolution limits of the current method.\\
In Fig.~\ref{FigureTension}a blue circles mark the tension measured as a function of temperature for adhering GUVs. In Fig.~\ref{FigureTension}b we show examples of power spectra fitted by Eq.~\ref{eqn:fitspectrum}. The tension typically lies in the interval $2\times 10^{-7}-2\times10^{-6}$ N~m$^{-1}$, with clear variations between different GUVs. In the tested range, the tension consistently displays a weak dependence on temperature changes.\\
For comparison, the membrane tension is measured on non-adhering GUVs supported by a passivated glass substrate. The values of $\sigma$ measured for non-adhering GUVs (red lozenges in Fig.~\ref{FigureTension}a) are significantly lower than those measured on adhering vesicles, falling within the range $10^{-9}-10^{-6}$ N~m$^{-1}$, and being clustered around  $10^{-8}$ N~m$^{-1}$. Furthermore, membrane tension of non-adhering GUVs typically displays a strong decrease upon increasing temperature. The large variability observed in the tension of  adhering and, in particular, of free GUVs, is ascribed to the polydispersity of electroformed samples, which produces vesicle populations with very different excess areas ($T_0$).\\
With the present technique we cannot access the tension of vesicles adhering to SBL for the case of higher DNA concentrations. Indeed, for values of $\sigma$ in the grey-shaded region on Fig.~\ref{FigureTension}a, the relevant portions of the fluctuation power spectra are masked by experimental noise deriving from the finite resolution of the contour-tracking procedure.\\
The membrane tension can be evaluated within the framework of our model. At equilibrium, the derivative of the interaction energy in Eq.~\ref{U1}, taken with respect to the GUV area, is
\begin{equation}\label{eqn:tensionDNA}
\frac{\partial U}{\partial A}=\sigma+\frac{\partial U_\mathrm{DNA}}{\partial A}=0,
\end{equation}
where we used
\begin{equation}
\sigma = \frac{\partial U_\mathrm{stretching}}{\partial A} = K_\mathrm{a} \frac{A-\tilde{A}}{\tilde{A}}.
\end{equation}
Equation \ref{eqn:tensionDNA} suggests that by measuring $\sigma$ we can directly probe the DNA-mediated forces. The blue-shaded region in Fig.~\ref{FigureTension}a marks the model predictions for $\sigma$, calculated using the parameters Table~\ref{Table}, $\rho_\mathrm{DNA}^\mathrm{low}=39\pm9$ $\mu$m$^{-2}$, and values of the neutral temperature covering the experimentally observed range ($-60\le T_0 \le 0^\circ$C). The size of the vesicles does not impact the predictions of $\sigma$, therefore we fix $R_\mathrm{0}=10$ $\mu$m. Solid blue lines mark the uncertainty interval propagating from the errorbars of the model parameters (Table~\ref{Table}). With no fitting parameters, we observe a semi-quantitative agreement between theory and experiments. In particular, the theory predicts the weak temperature-dependence of $\sigma$ observed in the experiments. The slight underestimation of the theoretical predictions as compared to the measured tension could be explained by an underestimation of the DNA-coating density. This evidence is consistent with the underestimation of the DNA melting temperature discussed above.\\
The green-dashed region in Fig.~\ref{FigureTension}a indicates the theoretical prediction calculated using $\rho_\mathrm{DNA}^\mathrm{med}=390\pm90$ $\mu$m$^{-2}$. The predicted tension falls within the non-accessible region.\\


\section{Conclusion}
In this article we experimentally investigate temperature-dependent adhesion of Giant-Unilamellar-Vesicles on supported lipid bilayers mediated by mobile DNA linkers. 
The simple geometry of the problem allows for an accurate characterisation of the morphology of adhering GUVs and the temperature dependent fraction of bound DNA tethers by means of 
confocal microscopy. For the first time to our knowledge, we quantify the temperature-dependent membrane tension induced by DNA bonds by analysing the thermal fluctuations of the GUVs imaged across their equatorial plane.\\
The experimental results are compared to theoretical predictions from our recently developed model,\cite{Parolini_NatComms_2015} which we here extend to the case of vesicle-plane adhesion. The model takes into account both the elastic deformation of the GUV and the statistical-mechanical details of the DNA-mediated interactions.\\
For sufficiently high DNA coverage, the adhesion contact angle exhibits a re-entrant temperature dependence. Upon heating from low temperature the contact angle increases, reaching a maximum at $T\simeq30-40^\circ$C. Upon further temperature increase, the contact angle drops. The re-entrance is less pronounced or absent for lower DNA coverage.
With a single fitting parameter, the model is capable of quantitatively predicting the low temperature regime and ascribes the increase in contact angle to the interplay between the temeperature-dependent excess area of the GUV and the entropic coupling between the hybridisation free-energy of the mobile tethers and the adhesion area. The theory is developed in the limit of strong adhesion, therefore it fails to predict the re-entrant behaviour of the adhesion area, caused by the weakening of the DNA bonds. The less-pronounced re-entrance observed for low DNA concentrations is ascribable to non-specific adhesive interactions that kick-in at high temperature, and are suppressed by steric repulsion for samples with high DNA coverage.\\
The melting of DNA bonds is investigated in-situ by FRET measurements. We observe a broad melting transition and find that bonds formed within the GUV-plane adhesion patch are more stable than in-plane bonds formed on free bilayers. With no fitting parameters our model can semi-quantitatively reproduce these features, although an underestimation of the melting temperature is observed.\\
Membrane tension measurements performed on adhering GUVs demonstrate a weak temperature dependence. In a similar range of temperatures, non-adhering GUVs exhibit significantly lower tension, rapidly decreasing upon heating. The differences in magnitude and trend demonstrates the role played by DNA in mediating membrane adhesion.
Experimental results are in semi-quantitative agreement with theoretical prediction, which further demonstrates the accuracy of the model used to describe hybridisation free-energy of tethered mobile linkers, and in particular the translational-entropic contributions that couples it to the adhesion area.\\

Our experimental observations and the agreement with the theoretical predictions help to clarify the complex mechanisms controlling adhesion of soft units mediated by multiple linkers. Besides the fundamental interest for the still poorly understood physics of multivalent interactions, our findings can help the design of functional, responsive, tissue-like materials with promised applications in biosensing, encapsulation-release mechanisms, and filtration. Finally, the conclusions drawn for our model system can be adapted to the quantitative description of cell adhesion and spreading on solid substrates, with possible biomedical implications in prosthetics and scaffoldings for tissue regeneration.\cite{Dalva_NatureRev_2007,Parsons_NatureRev_2010,Alberts,Bao_NMat_2003,Discher_Science_2005}

\subsection*{Acknowledgements} 
The authors acknowledge financial support from the EPSRC Programme Grant CAPITALS number EP/J017566/1 (PC, LP, LDM), the Ernest Oppenheimer Fund and Emmanuel College Cambridge (LDM), Universit\'{e} libre de Bruxelles (BMM), and Grant-in-Aid for JSPS Fellows Grant (No. 25-1270) (SFS), Project SPINNER 2013, Regione Emilia-Romagna (It), European Social Fund (DO). In compliance with our funders' requirements all the data underlying this article are available trough the corresponding authors.\\

\begin{appendix}
\section{Details on model derivation}

\subsection{Geometrical expressions}\label{App_geometry}
In the limit of strong adhesion the GUV will take the shape of a truncated sphere with contact angle $\theta$ (Fig.~\ref{Figure1}), which we take as the independent variable of our model.\cite{Ramachandran_Langmuir_2010} In this simple geometry, the contact (patch) area, total area, and volume of the GUV are respectively
\begin{eqnarray}
&A_\mathrm{p}=\pi R^2 \sin^2 {\theta} \label{eqn:Ap} \\
&A=\pi R^2 \left(1+\cos {\theta}\right)\left(3-\cos {\theta}\right)  \label{eqn:Area} \\
&V=\frac{\pi R^3}{3}\left(1+\cos {\theta}\right)^2\left(2-\cos {\theta} \right)  \label{eqn:V}.
\end{eqnarray}
In the limit of water-impermeable membranes, the internal volume of the GUVs can be taken as a constant
\begin{equation}\label{ConstantVolume}
V=\frac{4}{3}\pi R_0^3,
\end{equation}
where we introduce a reference radius $R_0$. By using Eqs.~\ref{eqn:V} and \ref{ConstantVolume}  we obtain
\begin{equation}\label{eqn:RofTheta}
R=R_0\left[\frac{4}{\left(1+\cos {\theta}\right)^2\left(2-\cos {\theta}\right)}\right]^{1/3},
\end{equation}
which can be inserted into Eqs. \ref{eqn:Ap} and \ref{eqn:Area} to make the $\theta$-dependence of $A_\mathrm{p}$  and $A$  explicit.\\
Let us now recall the definition of reduced volume of a vesicle\cite{Tordeux_PRE_2002,Ramachandran_Langmuir_2010}
\begin{equation}\label{eqn:reducedvolume}
v=\frac{\frac{3V}{4\pi}}{\left(\frac{\tilde{A}}{4\pi}\right)^{3/2}}.
\end{equation}
By combining Eq.~\ref{eqn:reducedvolume} with expression for the temperature-dependent unstretched area in Eq.~\ref{eqn:unstretched}, we obtain
\begin{equation}
v=\left[1+\alpha(T-T_0)\right]^{-3/2}.
\end{equation}
The reference temperature $T_0$ is therefore defined as the temperature at which a GUV has reduced volume equal to 1. For $T<T_0$, when $v>1$, an isolated vesicle resembles a turgid sphere, with non-zero membrane tension whereas for $T>T_0$, $v<1$, it assumes the features of a ``floppy" balloon, with excess area. At $T=T_0$ an isolated vesicle is a  perfect sphere with zero-membrane tension and radius equal to the reference radius $R_0$.\\
By combining Eqs.~\ref{eqn:Area}, \ref{eqn:RofTheta} and \ref{eqn:unstretched} we obtain an explicit, $\theta$-dependent expression for the stretching energy in Eq.~\ref{U2}. Note that Eq.~\ref{eqn:RofTheta} has been derived under a constant-volume assumption (Eq.~\ref{ConstantVolume}). Alternatively, an equivalent relation can be derived for water permeable -- solute impermeable -- GUVs, in which the volume is set by the balance between the osmotic pressure drop across the membrane and the Laplace pressure.\cite{Ramachandran_Langmuir_2010} In relevant experimental conditions the two assumptions lead to very similar results.\cite{Parolini_NatComms_2015}\\

\subsection{Free energy for bridge and loop formation.~~}\label{App_FreeEnergy}
For the case of mobile linkers, the configurational entropic contribution to the bridge/loop formation free energy $\Delta S^\mathrm{conf}_{\mathrm{b}/\ell}$ (Eq.~\ref{eqn:singleenergy}) can be split into a rotational and translational contribution
\begin{equation}\label{eqn:confent_app}
\Delta S_{\mathrm{b}/\ell}^\mathrm{conf}=\Delta S^\mathrm{rot} +\Delta S_{\mathrm{b}/\ell}^\mathrm{trans}.
\end{equation}
The rotational contribution takes the same expression for loop and bridge formation, and encodes for the reduction of configurational entropy following the hybridisation of two rigid tethers with fixed grafting sites\cite{Leunissen_JCP_2011,Parolini_NatComms_2015} 
\begin{equation}
\Delta S^\mathrm{rot} =k_\mathrm{B}\log\left[\frac{1}{4 \pi \rho_0 L^3}\right],\label{eqn:rotent}
\end{equation}
 where $\rho_0=1$M is a reference concentration.
The translational contribution encodes for the lateral confinement following the binding of two mobile tethers. For the case of loops, upon binding, two tethers initially capable of exploring the entire GUV surface area $A$, are confined to within a region $\sim L^2$ from each other\cite{Parolini_NatComms_2015}
\begin{equation}
\Delta S^\mathrm{trans}_\ell=k_\mathrm{B}\log\left[\frac{4\pi L^2 A}{A^2}\right]=k_\mathrm{B}\log\left[\frac{4\pi L^2}{A}\right]\label{eqn:transentLoops}.
\end{equation}
For the case of bridge formation, we consider a free linker on  the SBL, initially located within the contact area $A_\mathrm{p}$, and a second linker on the GUV, which is free to explore the entire surface area $A$. Upon binding, the area available to the pair is reduced to $4\pi L^2  A_\mathrm{p}$, resulting in
\begin{equation}
\Delta S^\mathrm{trans}_\mathrm{b}=k_\mathrm{B}\log\left[\frac{4\pi L^2 A_\mathrm{p}}{A A_\mathrm{p}}\right]=k_\mathrm{B}\log\left[\frac{4\pi L^2}{A}\right]\label{eqn:transentBridges}.
\end{equation}
We notice that, contrary to the case of two adhering GUVs,\cite{Parolini_NatComms_2015} here $\Delta S^\mathrm{trans}_\mathrm{b}=\Delta S^\mathrm{trans}_\ell$. By combining Eqs.~\ref{eqn:confent_app}--\ref{eqn:transentBridges} with Eq.~\ref{eqn:singleenergy}, we obtain the hybridisation free energy of bridge formation on the loops formation on the GUV in Eq.~\ref{DeltaG}.

\subsection{Fraction of loops and free tethers on the SBL.}\label{App_SBL}
We now focus on the description of the tethers anchored to the SBL, and calculate the equilibrium fraction of formed loops in the absence of an adhering GUV. This information is needed for the calculation of the GUV-SBL adhesive interaction as well as for a direct comparison with experimental data.\\
Let us consider a finite portion of the SBL of area $\Sigma$, containing two populations of $N$ linkers with $a$ and $a'$ sticky ends. Following Eq.~\ref{DeltaG}, the free energy for loop ($\ell$) formation on the SBL can be written as 
\begin{equation}\label{DeltaGLoopsSBL}
\Delta G_\ell^\mathrm{SBL} =\Delta G^0 - k_\mathrm{B}T\log\left[\frac{1}{\rho_0 L \Sigma}\right].
\end{equation}
By indicating as $N_\ell$ the number of loops within the SBL, and taking into account combinatorics, we can write the partition function of this systems as\cite{Parolini_NatComms_2015}
\begin{equation}
Z=\sum_{N_\ell} {N\choose N_\ell}^2 N_\ell ! \exp \left(-\beta N_\ell \Delta G_\ell^\mathrm{SBL}\right),
\end{equation}
which can be rearranged as
\begin{equation}\label{ZSBL}
Z=\sum_{N_\ell} e^{-S(N_\ell)}.
\end{equation}
We now consider the limit of an infinite SBL with a constant DNA surface density $c_0$, i.e. we take $N,\Sigma \rightarrow \infty$, with $c_0=N/\Sigma=\rho_\mathrm{DNA}/2$. By using the Stirling approximation we obtain
\begin{multline}
S(c_\ell) =\mathrm{const}\cdot \Sigma [-c_\ell \log c_\ell -2 (c_0-c_\ell)\log(c_0-c_\ell)\\
-c_\ell \beta \Delta G^0-c_\ell \log (\rho_0L) - c_\ell +\mathrm{const}],
\end{multline}
where we define the density of loops as $c_\ell=N_\ell/\Sigma$. Within the saddle-point approximation,\cite{Parolini_NatComms_2015} the sum in Eq. \ref{ZSBL} is dominated by the stationary point of $S$
\begin{equation}\label{saddleSBL}
\frac{\partial S}{\partial c_\ell}=0.
\end{equation}
By solving Eq.~\ref{saddleSBL} we obtain the expression in Eq.~\ref{eqn:freeSBL} in the text,
where we introduce the concentration of free tethers on the SBL $c_\mathrm{f}=c_0-c_\ell$. Note that with Eq.~\ref{eqn:freeSBL} we recover a simple mass-balance relation between loops and free tethers on the SBL, which ultimately results in
\begin{equation}\label{fractionofloopsSBL}
c_\ell=c_0\frac{2 q_\mathrm{SBL} + 1 - \sqrt{4 q_\mathrm{SBL}+1}}{2 q_\mathrm{SBL}},
\end{equation}
where
\begin{equation}\label{qloopsSBL}
q_\mathrm{SBL}=\frac{c_0}{\rho_0 L} \exp[-\beta \Delta G_0].
\end{equation}\\
The fraction of loops formed within the bilayer is $x_\ell^\mathrm{SBL}=c_\ell/c_0$.

\subsection{Combinatorial effects.~~}\label{App_Comb}
Given the expressions for the hybridisation free-energy of a single bridge/loop (Eq.~\ref{DeltaG}), a combinatorial approach is required to compute the overall DNA-mediated interaction energy.\\
Following the derivation carried out to describe loop formation on the SBL, we indicate the total number of tethers with $a$ ($a'$) sticky ends on the GUV as $N$, and define $N_\ell$ as the number of those tethers linked in loops. We indicate as $N_{\mathrm{b}i}$, with $i=1,2$, the number of tethers forming bridges with those on the SBL, with the index $i$ referring to $a$ and $a'$ sticky ends. We label as $N_{\mathrm{f}i}$ the number of free tethers on the SBL located within the adhesion patch, where the index $i=1,2$ now refers to $a'$ and $a$ sticky ends. The partition function of the system of linkers can be written as
\begin{equation}\label{eqn:z1}
\begin{split}
 z(N_{\mathrm{f}1},N_{\mathrm{f}1},N) \nonumber = \sum_{N_{\mathrm{b}i},N_\ell} \bigl[ \Omega_{N_{\mathrm{f}1},N_{\mathrm{f}2},N}(N_{\mathrm{b}1},N_{\mathrm{b}1},N_\ell)\\
\exp \left[-\beta (N_\mathrm{b1} + N_\mathrm{b2} +N_\ell) \Delta G\right]\bigr],
\end{split}
\end{equation}
where the number of possible configurations for a given $N_\ell$ and $N_{\mathrm{b}i}$ is
\begin{multline}
\Omega_{N_{\mathrm{f}1},N_{\mathrm{f}2},N}(N_{\mathrm{b}1},N_{\mathrm{b}1},N_\ell)= \\ N_\ell!\prod_{i=1,2} {N_{\mathrm{f}i} \choose N_{\mathrm{b}i}}{N \choose N_{\mathrm{b}i}} { {N-N_{\mathrm{b}i}} \choose N_\ell}N_{\mathrm{b}i}!
\end{multline}
To account for strand-concentration fluctuations within the adhesion patch, we need to consider that $N_{\mathrm{f}i}$ is Poisson-distributed around its average value $\bar{N}_{\mathrm{f}}$
\begin{equation}\label{enq:Poisson}
P(N_{\mathrm{f}i},\bar{N}_\mathrm{f})=\exp[-\bar{N}_\mathrm{f}]\frac{\bar{N}_\mathrm{f}^{N_{\mathrm{f}i}}}{N_{\mathrm{f}i} !}.
\end{equation}
Using Eq.~\ref{fractionofloopsSBL}, and recalling that $c_\mathrm{f}=c_0-c_\ell$ is the concentration of free tethers within the SBL, we find
\begin{equation}\label{Nbar}
\bar{N}_{\mathrm{f}}(\theta)=c_\mathrm{f} A_\mathrm{p}(\theta),
\end{equation}
where we highlighted the strong dependence on the contact angle $\theta$.
Using Eqs. \ref{eqn:z1} and \ref{enq:Poisson} we write the full partition function as
\begin{equation}\label{eqn:z2}
Z_{N_{\mathrm{f}1},N_{\mathrm{f}2}}(\bar{N}_\mathrm{f},N)=\sum_{N_\mathrm{f}} P(N_{\mathrm{f}1},\bar{N}_\mathrm{f}) P(N_{\mathrm{f}2},\bar{N}_\mathrm{f}) z(N_{\mathrm{f}1},N_{\mathrm{f}1},N).
\end{equation}
Equation \ref{eqn:z2} can be rearranged as
\begin{equation}\label{eqn:z3}
Z_{N_{\mathrm{f}1},N_{\mathrm{f}2}}(\bar{N}_\mathrm{f},N)=\sum_{N_{\mathrm{f}1},N_{\mathrm{f}2},N_\ell,N_{\mathrm{b}1},N_{\mathrm{b}2}} \exp[-N  \mathcal{A}].
\end{equation}
By defining the fractions $x_y=N_\mathrm{y}/N$ ($y=$b1, b2, l, f1, f2), and using the Stirling approximation, $\mathcal{A}$ can be expressed as
\begin{multline}\label{eqn:A}
 \mathcal{A}=   \beta \Delta G^*x_\ell  + x_{\ell}\left(\log  x_{\ell} +1\right) + \\
 \sum_{i=1,2}  \Bigl[ \beta \Delta G^*x_{\mathrm{b}i} + x_{\mathrm{f}i} \left( \log N - \log \bar{N}_\mathrm{f} -1 \right) + \\
 x_{\mathrm{b}i} \left(\log  x_{\mathrm{b}i} +1\right)  +\left(x_{\mathrm{f}i}-x_{\mathrm{b}i}\right)\log \left(x_{\mathrm{f}i}-x_{\mathrm{b}i}\right)\\
+ \left(1-x_{\ell}-x_{\mathrm{b}i}\right) \log  \left(1-x_{\ell}-x_{\mathrm{b}i}\right) \Bigr].
\end{multline}
Note that in Eq.~\ref{eqn:A} we re-defined the hybridisation free energy for bridge and loop formation as
\begin{equation}\label{eqnDGstar}
\Delta G^* = \Delta G - k_\mathrm{B}T \log N.
\end{equation}
For typical experimental conditions, the attractive combinatorial term $-k_\mathrm{B}T \log N$ in Eq. \ref{eqnDGstar} can be estimated in $\approx-10 k_\mathrm{B}T$.\cite{Parolini_NatComms_2015} Within the saddle-point approximation, the sum in Eq.~\ref{eqn:z3} is dominated by the stationary point of $\mathcal{A}$
\begin{equation}\label{SaddlePoint}
\frac{\partial \mathcal{A}}{\partial x_y} = 0 \mbox{ with $y$=b1, b2, l, f1, f2}.
\end{equation}
From the saddle-point equations Eq.~\ref{SaddlePoint} we obtain Eqs.~\ref{BridgesImpl} and \ref{LoopsImpl} in the text, where we find $x_{\mathrm{b}1}=x_{\mathrm{b}2}=x_\mathrm{b}$. 
By solving Eqs.~\ref{BridgesImpl} and ~\ref{LoopsImpl} we obtain
\begin{equation}
x_\mathrm{b}=\frac{q_\mathrm{b}\left(\sqrt{q_\mathrm{b}^2+2q_\mathrm{b}+4q_\ell +1}- q_\mathrm{b}- 1 \right) }{2q_\ell} \label{Bridges}\\ 
\end{equation}
\begin{equation}
x_\ell=\frac{ q_\mathrm{b}^2+2q_\mathrm{b}+2q_\ell +1 -\left(q_\mathrm{b}+1\right) \sqrt{q_\mathrm{b}^2+2q_\mathrm{b}+4q_\ell +1}}{2q_\ell}. \label{Loops}
\end{equation}
Note that for simplicity the fraction of bridges and loops are indicated as $x_{\mathrm{b}/\ell}$.
The saddle point equations for $x_\mathrm{f}$ ($x_{\mathrm{f},1}=x_{\mathrm{f},2}$) read $x_\mathrm{f}-x_\mathrm{b}=\bar{N_\mathrm{f}}/N$, which confirms that the density of the free tethers in in the patch region is equal to that of the reservoir, as expected.\\
By inserting the saddle-point solutions for $x_\ell,$ $x_\mathrm{b}$, and $x_\mathrm{f}$ in Eqs.~\ref{eqn:z3} and \ref{eqn:A} we can calculate the free energy $U_\mathrm{hyb}$ (Eq.~\ref{UHyb}).\cite{Angioletti-Uberti_JCP_2013}

\subsection{Reference energy.}\label{App_Ref}
The reference energy $U_0$ in Eq.~\ref{U1} is calculated for isolated GUV and SBL and can be written as
\begin{equation}\label{U0}
U_0=U_{0}^\mathrm{stretching}+U_{0}^\mathrm{DNA}.
\end{equation}
The stretching term is\cite{Parolini_NatComms_2015}
\begin{equation}\label{UST0}
U_0^\mathrm{stretching}=
\begin{cases}
0 \mbox{ if } T \ge T_0\\
K_\mathrm{a}\frac{\left(A_0-\tilde{A}\right)^2}{A_0} \mbox{ if } T< T_0.
\end{cases}
\end{equation}
Note that the stretching contribution is only present for pre-stretched vesicles, i.e. if the reduced volume is $v>1$ (i.e. $T<T_0$).~\cite{Ramachandran_Langmuir_2010}
The DNA contribution is calculated for a GUV of area equal to the unstretched area $\tilde{A}$, in which only loops can form. By following the steps outlined in section \ref{App_Comb} and in ref.\cite{Parolini_NatComms_2015} we calculate the fraction of tethers involved in loops
\begin{equation}
{x}^0_\ell=\frac{ 2 q_\ell +1  - \sqrt{4 q_\ell+1}}{2 q_\ell},
\end{equation}
where $q_\ell$ is given by Eq.~\ref{QLoops}. The DNA part of the reference energy is 
\begin{equation}\label{UDNA0}
U_0^\mathrm{DNA}=N k_\mathrm{B} T \left[x^0_\ell + 2 \log \left(1-x^0_\ell\right)-2\frac{\tilde{N}_\mathrm{f}}{N}-2 \log\left(\frac{\tilde{A}}{A_0}\right)\right],
\end{equation}
where $\tilde{N}_\mathrm{f}=c_\mathrm{f}\tilde{A}_\mathrm{p}$ is the number of free tethers present within area $\tilde{A}_\mathrm{p}$ on the SBL. $\tilde{A}_\mathrm{p}$ is the zero-stretching adhesion area, which the GUV-SBL system would form for negligibly small attractive forces when $T>T_0$, as derived in ref.\cite{Parolini_NatComms_2015}\\
Note that $U_0$ does not depend on the contact angle $\theta$ therefore its form does not influence the equilibrium features of the system.
\end{appendix}

\bibliography{DNAVesiclesBIB} 

\begin{thebibliography}{62}%
\makeatletter
\providecommand \@ifxundefined [1]{%
 \@ifx{#1\undefined}
}%
\providecommand \@ifnum [1]{%
 \ifnum #1\expandafter \@firstoftwo
 \else \expandafter \@secondoftwo
 \fi
}%
\providecommand \@ifx [1]{%
 \ifx #1\expandafter \@firstoftwo
 \else \expandafter \@secondoftwo
 \fi
}%
\providecommand \natexlab [1]{#1}%
\providecommand \enquote  [1]{``#1''}%
\providecommand \bibnamefont  [1]{#1}%
\providecommand \bibfnamefont [1]{#1}%
\providecommand \citenamefont [1]{#1}%
\providecommand \href@noop [0]{\@secondoftwo}%
\providecommand \href [0]{\begingroup \@sanitize@url \@href}%
\providecommand \@href[1]{\@@startlink{#1}\@@href}%
\providecommand \@@href[1]{\endgroup#1\@@endlink}%
\providecommand \@sanitize@url [0]{\catcode `\\12\catcode `\$12\catcode
  `\&12\catcode `\#12\catcode `\^12\catcode `\_12\catcode `\%12\relax}%
\providecommand \@@startlink[1]{}%
\providecommand \@@endlink[0]{}%
\providecommand \url  [0]{\begingroup\@sanitize@url \@url }%
\providecommand \@url [1]{\endgroup\@href {#1}{\urlprefix }}%
\providecommand \urlprefix  [0]{URL }%
\providecommand \Eprint [0]{\href }%
\providecommand \doibase [0]{http://dx.doi.org/}%
\providecommand \selectlanguage [0]{\@gobble}%
\providecommand \bibinfo  [0]{\@secondoftwo}%
\providecommand \bibfield  [0]{\@secondoftwo}%
\providecommand \translation [1]{[#1]}%
\providecommand \BibitemOpen [0]{}%
\providecommand \bibitemStop [0]{}%
\providecommand \bibitemNoStop [0]{.\EOS\space}%
\providecommand \EOS [0]{\spacefactor3000\relax}%
\providecommand \BibitemShut  [1]{\csname bibitem#1\endcsname}%
\let\auto@bib@innerbib\@empty
\bibitem [{\citenamefont {Di~Michele}\ and\ \citenamefont
  {Eiser}(2013)}]{Di-Michele_PCCP_2013}%
  \BibitemOpen
  \bibfield  {author} {\bibinfo {author} {\bibfnamefont {L.}~\bibnamefont
  {Di~Michele}}\ and\ \bibinfo {author} {\bibfnamefont {E.}~\bibnamefont
  {Eiser}},\ }\href@noop {} {\bibfield  {journal} {\bibinfo  {journal} {Phys.
  Chem. Chem. Phys.}\ }\textbf {\bibinfo {volume} {15}},\ \bibinfo {pages}
  {3115} (\bibinfo {year} {2013})}\BibitemShut {NoStop}%
\bibitem [{\citenamefont {Mirkin}\ \emph {et~al.}(1996)\citenamefont {Mirkin},
  \citenamefont {Letsinger}, \citenamefont {Mucic},\ and\ \citenamefont
  {Storhoff}}]{Mirkin_Nature_1996}%
  \BibitemOpen
  \bibfield  {author} {\bibinfo {author} {\bibfnamefont {C.~A.}\ \bibnamefont
  {Mirkin}}, \bibinfo {author} {\bibfnamefont {R.~L.}\ \bibnamefont
  {Letsinger}}, \bibinfo {author} {\bibfnamefont {R.~C.}\ \bibnamefont
  {Mucic}}, \ and\ \bibinfo {author} {\bibfnamefont {J.~J.}\ \bibnamefont
  {Storhoff}},\ }\href@noop {} {\bibfield  {journal} {\bibinfo  {journal}
  {Nature}\ }\textbf {\bibinfo {volume} {382}},\ \bibinfo {pages} {607}
  (\bibinfo {year} {1996})}\BibitemShut {NoStop}%
\bibitem [{\citenamefont {Alivisatos}\ \emph {et~al.}(1996)\citenamefont
  {Alivisatos}, \citenamefont {Johnsson}, \citenamefont {Peng}, \citenamefont
  {Wilson}, \citenamefont {Loweth}, \citenamefont {Bruchez},\ and\
  \citenamefont {Schultz}}]{Alivisatos_Nature_1996}%
  \BibitemOpen
  \bibfield  {author} {\bibinfo {author} {\bibfnamefont {A.~P.}\ \bibnamefont
  {Alivisatos}}, \bibinfo {author} {\bibfnamefont {K.~P.}\ \bibnamefont
  {Johnsson}}, \bibinfo {author} {\bibfnamefont {X.}~\bibnamefont {Peng}},
  \bibinfo {author} {\bibfnamefont {T.~E.}\ \bibnamefont {Wilson}}, \bibinfo
  {author} {\bibfnamefont {C.~J.}\ \bibnamefont {Loweth}}, \bibinfo {author}
  {\bibfnamefont {M.~P.}\ \bibnamefont {Bruchez}}, \ and\ \bibinfo {author}
  {\bibfnamefont {P.~G.}\ \bibnamefont {Schultz}},\ }\href@noop {} {\bibfield
  {journal} {\bibinfo  {journal} {Nature}\ }\textbf {\bibinfo {volume} {382}},\
  \bibinfo {pages} {609} (\bibinfo {year} {1996})}\BibitemShut {NoStop}%
\bibitem [{\citenamefont {Hill}\ \emph {et~al.}(2008)\citenamefont {Hill},
  \citenamefont {Macfarlane}, \citenamefont {Senesi}, \citenamefont {Lee},
  \citenamefont {Park},\ and\ \citenamefont {Mirkin}}]{Hill_NanoLetters_2008}%
  \BibitemOpen
  \bibfield  {author} {\bibinfo {author} {\bibfnamefont {H.~D.}\ \bibnamefont
  {Hill}}, \bibinfo {author} {\bibfnamefont {R.~J.}\ \bibnamefont
  {Macfarlane}}, \bibinfo {author} {\bibfnamefont {A.~J.}\ \bibnamefont
  {Senesi}}, \bibinfo {author} {\bibfnamefont {B.}~\bibnamefont {Lee}},
  \bibinfo {author} {\bibfnamefont {S.~Y.}\ \bibnamefont {Park}}, \ and\
  \bibinfo {author} {\bibfnamefont {C.~A.}\ \bibnamefont {Mirkin}},\ }\href
  {\doibase 10.1021/nl8011787} {\bibfield  {journal} {\bibinfo  {journal} {Nano
  Letters}\ }\textbf {\bibinfo {volume} {8}},\ \bibinfo {pages} {2341}
  (\bibinfo {year} {2008})}\BibitemShut {NoStop}%
\bibitem [{\citenamefont {Maye}\ \emph {et~al.}(2010)\citenamefont {Maye},
  \citenamefont {Kumara}, \citenamefont {Nykypanchuk}, \citenamefont
  {Sherman},\ and\ \citenamefont {Gang}}]{Maye_NatNano_2010}%
  \BibitemOpen
  \bibfield  {author} {\bibinfo {author} {\bibfnamefont {M.~M.}\ \bibnamefont
  {Maye}}, \bibinfo {author} {\bibfnamefont {M.~T.}\ \bibnamefont {Kumara}},
  \bibinfo {author} {\bibfnamefont {D.}~\bibnamefont {Nykypanchuk}}, \bibinfo
  {author} {\bibfnamefont {W.~B.}\ \bibnamefont {Sherman}}, \ and\ \bibinfo
  {author} {\bibfnamefont {O.}~\bibnamefont {Gang}},\ }\href@noop {} {\bibfield
   {journal} {\bibinfo  {journal} {Nat. Nano.}\ }\textbf {\bibinfo {volume}
  {5}},\ \bibinfo {pages} {116} (\bibinfo {year} {2010})}\BibitemShut {NoStop}%
\bibitem [{\citenamefont {Nykypanchuk}\ \emph {et~al.}(2008)\citenamefont
  {Nykypanchuk}, \citenamefont {Maye}, \citenamefont {van~der Lelie},\ and\
  \citenamefont {Gang}}]{Nykypanchuk_Nature_2008}%
  \BibitemOpen
  \bibfield  {author} {\bibinfo {author} {\bibfnamefont {D.}~\bibnamefont
  {Nykypanchuk}}, \bibinfo {author} {\bibfnamefont {M.~M.}\ \bibnamefont
  {Maye}}, \bibinfo {author} {\bibfnamefont {D.}~\bibnamefont {van~der Lelie}},
  \ and\ \bibinfo {author} {\bibfnamefont {O.}~\bibnamefont {Gang}},\
  }\href@noop {} {\bibfield  {journal} {\bibinfo  {journal} {Nature}\ }\textbf
  {\bibinfo {volume} {451}},\ \bibinfo {pages} {549} (\bibinfo {year}
  {2008})}\BibitemShut {NoStop}%
\bibitem [{\citenamefont {Cigler}\ \emph {et~al.}(2010)\citenamefont {Cigler},
  \citenamefont {Lytton-Jean}, \citenamefont {Anderson}, \citenamefont {Finn},\
  and\ \citenamefont {Park}}]{Park_NMat_2010}%
  \BibitemOpen
  \bibfield  {author} {\bibinfo {author} {\bibfnamefont {P.}~\bibnamefont
  {Cigler}}, \bibinfo {author} {\bibfnamefont {A.~K.~R.}\ \bibnamefont
  {Lytton-Jean}}, \bibinfo {author} {\bibfnamefont {D.~G.}\ \bibnamefont
  {Anderson}}, \bibinfo {author} {\bibfnamefont {M.~G.}\ \bibnamefont {Finn}},
  \ and\ \bibinfo {author} {\bibfnamefont {S.~Y.}\ \bibnamefont {Park}},\
  }\href@noop {} {\bibfield  {journal} {\bibinfo  {journal} {Nat. Mater.}\
  }\textbf {\bibinfo {volume} {9}},\ \bibinfo {pages} {918} (\bibinfo {year}
  {2010})}\BibitemShut {NoStop}%
\bibitem [{\citenamefont {Macfarlane}\ \emph {et~al.}(2011)\citenamefont
  {Macfarlane}, \citenamefont {Lee}, \citenamefont {Jones}, \citenamefont
  {Harris}, \citenamefont {Schatz},\ and\ \citenamefont
  {Mirkin}}]{Mirkin_Science_2011}%
  \BibitemOpen
  \bibfield  {author} {\bibinfo {author} {\bibfnamefont {R.~J.}\ \bibnamefont
  {Macfarlane}}, \bibinfo {author} {\bibfnamefont {B.}~\bibnamefont {Lee}},
  \bibinfo {author} {\bibfnamefont {M.~R.}\ \bibnamefont {Jones}}, \bibinfo
  {author} {\bibfnamefont {N.}~\bibnamefont {Harris}}, \bibinfo {author}
  {\bibfnamefont {G.~C.}\ \bibnamefont {Schatz}}, \ and\ \bibinfo {author}
  {\bibfnamefont {C.~A.}\ \bibnamefont {Mirkin}},\ }\href@noop {} {\bibfield
  {journal} {\bibinfo  {journal} {Science}\ }\textbf {\bibinfo {volume}
  {334}},\ \bibinfo {pages} {204} (\bibinfo {year} {2011})}\BibitemShut
  {NoStop}%
\bibitem [{\citenamefont {Di~Michele}\ \emph {et~al.}(2013)\citenamefont
  {Di~Michele}, \citenamefont {Varrato}, \citenamefont {Kotar}, \citenamefont
  {Nathan}, \citenamefont {Foffi},\ and\ \citenamefont
  {Eiser}}]{Di-Michele_NatComm_2013}%
  \BibitemOpen
  \bibfield  {author} {\bibinfo {author} {\bibfnamefont {L.}~\bibnamefont
  {Di~Michele}}, \bibinfo {author} {\bibfnamefont {F.}~\bibnamefont {Varrato}},
  \bibinfo {author} {\bibfnamefont {J.}~\bibnamefont {Kotar}}, \bibinfo
  {author} {\bibfnamefont {S.~H.}\ \bibnamefont {Nathan}}, \bibinfo {author}
  {\bibfnamefont {G.}~\bibnamefont {Foffi}}, \ and\ \bibinfo {author}
  {\bibfnamefont {E.}~\bibnamefont {Eiser}},\ }\href@noop {} {\bibfield
  {journal} {\bibinfo  {journal} {Nat. Commun.}\ }\textbf {\bibinfo {volume}
  {4}},\ \bibinfo {pages} {2007} (\bibinfo {year} {2013})}\BibitemShut
  {NoStop}%
\bibitem [{\citenamefont {Varrato}\ \emph {et~al.}(2012)\citenamefont
  {Varrato}, \citenamefont {Di~Michele}, \citenamefont {Belushkin},
  \citenamefont {Dorsaz}, \citenamefont {Nathan}, \citenamefont {Eiser},\ and\
  \citenamefont {Foffi}}]{Varrato_PNAS_2012}%
  \BibitemOpen
  \bibfield  {author} {\bibinfo {author} {\bibfnamefont {F.}~\bibnamefont
  {Varrato}}, \bibinfo {author} {\bibfnamefont {L.}~\bibnamefont {Di~Michele}},
  \bibinfo {author} {\bibfnamefont {M.}~\bibnamefont {Belushkin}}, \bibinfo
  {author} {\bibfnamefont {N.}~\bibnamefont {Dorsaz}}, \bibinfo {author}
  {\bibfnamefont {S.~H.}\ \bibnamefont {Nathan}}, \bibinfo {author}
  {\bibfnamefont {E.}~\bibnamefont {Eiser}}, \ and\ \bibinfo {author}
  {\bibfnamefont {G.}~\bibnamefont {Foffi}},\ }\href@noop {} {\bibfield
  {journal} {\bibinfo  {journal} {Proc. Natl. Acad. Sci. U.S.A.}\ }\textbf
  {\bibinfo {volume} {109}},\ \bibinfo {pages} {19155} (\bibinfo {year}
  {2012})}\BibitemShut {NoStop}%
\bibitem [{\citenamefont {Di~Michele}\ \emph
  {et~al.}(2014{\natexlab{a}})\citenamefont {Di~Michele}, \citenamefont
  {Fiocco}, \citenamefont {Varrato}, \citenamefont {Sastry}, \citenamefont
  {Eiser},\ and\ \citenamefont {Foffi}}]{Di-Michele_SoftMatter_2014}%
  \BibitemOpen
  \bibfield  {author} {\bibinfo {author} {\bibfnamefont {L.}~\bibnamefont
  {Di~Michele}}, \bibinfo {author} {\bibfnamefont {D.}~\bibnamefont {Fiocco}},
  \bibinfo {author} {\bibfnamefont {F.}~\bibnamefont {Varrato}}, \bibinfo
  {author} {\bibfnamefont {S.}~\bibnamefont {Sastry}}, \bibinfo {author}
  {\bibfnamefont {E.}~\bibnamefont {Eiser}}, \ and\ \bibinfo {author}
  {\bibfnamefont {G.}~\bibnamefont {Foffi}},\ }\href@noop {} {\bibfield
  {journal} {\bibinfo  {journal} {Soft Matter}\ }\textbf {\bibinfo {volume}
  {10}},\ \bibinfo {pages} {3633} (\bibinfo {year}
  {2014}{\natexlab{a}})}\BibitemShut {NoStop}%
\bibitem [{\citenamefont {Young}\ \emph {et~al.}(2014)\citenamefont {Young},
  \citenamefont {Ross}, \citenamefont {Blaber}, \citenamefont {Rycenga},
  \citenamefont {Jones}, \citenamefont {Zhang}, \citenamefont {Senesi},
  \citenamefont {Lee}, \citenamefont {Schatz},\ and\ \citenamefont
  {Mirkin}}]{Young_AdvMat_2014}%
  \BibitemOpen
  \bibfield  {author} {\bibinfo {author} {\bibfnamefont {K.~L.}\ \bibnamefont
  {Young}}, \bibinfo {author} {\bibfnamefont {M.~B.}\ \bibnamefont {Ross}},
  \bibinfo {author} {\bibfnamefont {M.~G.}\ \bibnamefont {Blaber}}, \bibinfo
  {author} {\bibfnamefont {M.}~\bibnamefont {Rycenga}}, \bibinfo {author}
  {\bibfnamefont {M.~R.}\ \bibnamefont {Jones}}, \bibinfo {author}
  {\bibfnamefont {C.}~\bibnamefont {Zhang}}, \bibinfo {author} {\bibfnamefont
  {A.~J.}\ \bibnamefont {Senesi}}, \bibinfo {author} {\bibfnamefont
  {B.}~\bibnamefont {Lee}}, \bibinfo {author} {\bibfnamefont {G.~C.}\
  \bibnamefont {Schatz}}, \ and\ \bibinfo {author} {\bibfnamefont {C.~A.}\
  \bibnamefont {Mirkin}},\ }\href {\doibase 10.1002/adma.201302938} {\bibfield
  {journal} {\bibinfo  {journal} {Advanced Materials}\ }\textbf {\bibinfo
  {volume} {26}},\ \bibinfo {pages} {653} (\bibinfo {year} {2014})}\BibitemShut
  {NoStop}%
\bibitem [{\citenamefont {Thaxton}\ \emph {et~al.}(2009)\citenamefont
  {Thaxton}, \citenamefont {Elghanian}, \citenamefont {Thomas}, \citenamefont
  {Stoeva}, \citenamefont {Lee}, \citenamefont {Smith}, \citenamefont
  {Schaeffer}, \citenamefont {Klocker}, \citenamefont {Horninger},
  \citenamefont {Bartsch},\ and\ \citenamefont {Mirkin}}]{Thaxton_PNAS_2009}%
  \BibitemOpen
  \bibfield  {author} {\bibinfo {author} {\bibfnamefont {C.~S.}\ \bibnamefont
  {Thaxton}}, \bibinfo {author} {\bibfnamefont {R.}~\bibnamefont {Elghanian}},
  \bibinfo {author} {\bibfnamefont {A.~D.}\ \bibnamefont {Thomas}}, \bibinfo
  {author} {\bibfnamefont {S.~I.}\ \bibnamefont {Stoeva}}, \bibinfo {author}
  {\bibfnamefont {J.-S.}\ \bibnamefont {Lee}}, \bibinfo {author} {\bibfnamefont
  {N.~D.}\ \bibnamefont {Smith}}, \bibinfo {author} {\bibfnamefont {A.~J.}\
  \bibnamefont {Schaeffer}}, \bibinfo {author} {\bibfnamefont {H.}~\bibnamefont
  {Klocker}}, \bibinfo {author} {\bibfnamefont {W.}~\bibnamefont {Horninger}},
  \bibinfo {author} {\bibfnamefont {G.}~\bibnamefont {Bartsch}}, \ and\
  \bibinfo {author} {\bibfnamefont {C.~A.}\ \bibnamefont {Mirkin}},\ }\href
  {http://www.pnas.org/content/106/44/18437.abstract N2 - We report the
  development of a previously undescribed gold nanoparticle bio-barcode assay
  probe for the detection of prostate specific antigen (PSA) at 330 fg/mL,
  automation of the assay, and the results of a clinical pilot study designed
  to assess the ability of the assay to detect PSA in the serum of 18 men who
  have undergone radical prostatectomy for prostate cancer. Due to a lack of
  sensitivity, available PSA immunoassays are often not capable of detecting
  PSA in the serum of men after radical prostatectomy. This new bio-barcode PSA
  assay is \UTF{2248}300 times more sensitive than commercial immunoassays.
  Significantly, with the barcode assay, every patient in this cohort had a
  measurable serum PSA level after radical prostatectomy. Patients were
  separated into categories based on PSA levels as a function of time. One
  group of patients showed low levels of PSA with no significant increase with
  time and did not recur. Others showed, at some point postprostatectomy,
  rising PSA levels. The majority recurred. Therefore, this new ultrasensitive
  assay points to significant possible outcomes: (i) The ability to tell
  patients, who have undetectable PSA levels with conventional assays, but
  detectable and nonrising levels with the barcode assay, that their cancer
  will not recur. (ii) The ability to assign recurrence earlier because of the
  ability to measure increasing levels of PSA before conventional tools can
  make such assignments. (iii) The ability to use PSA levels that are not
  detectable with conventional assays to follow the response of patients to
  adjuvant or salvage therapies.} {\bibfield  {journal} {\bibinfo  {journal}
  {Proceedings of the National Academy of Sciences}\ }\textbf {\bibinfo
  {volume} {106}},\ \bibinfo {pages} {18437} (\bibinfo {year}
  {2009})}\BibitemShut {NoStop}%
\bibitem [{\citenamefont {Wu}\ \emph {et~al.}(2014)\citenamefont {Wu},
  \citenamefont {Choi}, \citenamefont {Zhang}, \citenamefont {Hao},\ and\
  \citenamefont {Mirkin}}]{Wu_JACS_2014}%
  \BibitemOpen
  \bibfield  {author} {\bibinfo {author} {\bibfnamefont {X.~A.}\ \bibnamefont
  {Wu}}, \bibinfo {author} {\bibfnamefont {C.~H.~J.}\ \bibnamefont {Choi}},
  \bibinfo {author} {\bibfnamefont {C.}~\bibnamefont {Zhang}}, \bibinfo
  {author} {\bibfnamefont {L.}~\bibnamefont {Hao}}, \ and\ \bibinfo {author}
  {\bibfnamefont {C.~A.}\ \bibnamefont {Mirkin}},\ }\href@noop {} {\bibfield
  {journal} {\bibinfo  {journal} {J. Am. Chem. Soc.}\ }\textbf {\bibinfo
  {volume} {136}},\ \bibinfo {pages} {7726} (\bibinfo {year}
  {2014})}\BibitemShut {NoStop}%
\bibitem [{\citenamefont {Rosi}\ \emph {et~al.}(2006)\citenamefont {Rosi},
  \citenamefont {Giljohann}, \citenamefont {Thaxton}, \citenamefont
  {Lytton-Jean}, \citenamefont {Han},\ and\ \citenamefont
  {Mirkin}}]{Rosi_Science_2006}%
  \BibitemOpen
  \bibfield  {author} {\bibinfo {author} {\bibfnamefont {N.~L.}\ \bibnamefont
  {Rosi}}, \bibinfo {author} {\bibfnamefont {D.~A.}\ \bibnamefont {Giljohann}},
  \bibinfo {author} {\bibfnamefont {C.~S.}\ \bibnamefont {Thaxton}}, \bibinfo
  {author} {\bibfnamefont {A.~K.~R.}\ \bibnamefont {Lytton-Jean}}, \bibinfo
  {author} {\bibfnamefont {M.~S.}\ \bibnamefont {Han}}, \ and\ \bibinfo
  {author} {\bibfnamefont {C.~A.}\ \bibnamefont {Mirkin}},\ }\href@noop {}
  {\bibfield  {journal} {\bibinfo  {journal} {Science}\ }\textbf {\bibinfo
  {volume} {312}},\ \bibinfo {pages} {1027} (\bibinfo {year}
  {2006})}\BibitemShut {NoStop}%
\bibitem [{\citenamefont {Cutler}\ \emph {et~al.}(2012)\citenamefont {Cutler},
  \citenamefont {Auyeung},\ and\ \citenamefont {Mirkin}}]{Cutler_JACS_2012}%
  \BibitemOpen
  \bibfield  {author} {\bibinfo {author} {\bibfnamefont {J.~I.}\ \bibnamefont
  {Cutler}}, \bibinfo {author} {\bibfnamefont {E.}~\bibnamefont {Auyeung}}, \
  and\ \bibinfo {author} {\bibfnamefont {C.~A.}\ \bibnamefont {Mirkin}},\
  }\href@noop {} {\bibfield  {journal} {\bibinfo  {journal} {J. Am. Chem.
  Soc.}\ }\textbf {\bibinfo {volume} {134}},\ \bibinfo {pages} {1376} (\bibinfo
  {year} {2012})}\BibitemShut {NoStop}%
\bibitem [{\citenamefont {Dreyfus}\ \emph {et~al.}(2009)\citenamefont
  {Dreyfus}, \citenamefont {Leunissen}, \citenamefont {Sha}, \citenamefont
  {Tkachenko}, \citenamefont {Seeman}, \citenamefont {Pine},\ and\
  \citenamefont {Chaikin}}]{Dreyfus_PRL_2009}%
  \BibitemOpen
  \bibfield  {author} {\bibinfo {author} {\bibfnamefont {R.}~\bibnamefont
  {Dreyfus}}, \bibinfo {author} {\bibfnamefont {M.~E.}\ \bibnamefont
  {Leunissen}}, \bibinfo {author} {\bibfnamefont {R.}~\bibnamefont {Sha}},
  \bibinfo {author} {\bibfnamefont {A.~V.}\ \bibnamefont {Tkachenko}}, \bibinfo
  {author} {\bibfnamefont {N.~C.}\ \bibnamefont {Seeman}}, \bibinfo {author}
  {\bibfnamefont {D.~J.}\ \bibnamefont {Pine}}, \ and\ \bibinfo {author}
  {\bibfnamefont {P.~M.}\ \bibnamefont {Chaikin}},\ }\href@noop {} {\bibfield
  {journal} {\bibinfo  {journal} {Phys. Rev. Lett.}\ }\textbf {\bibinfo
  {volume} {102}},\ \bibinfo {pages} {048301} (\bibinfo {year}
  {2009})}\BibitemShut {NoStop}%
\bibitem [{\citenamefont {Varilly}\ \emph {et~al.}(2012)\citenamefont
  {Varilly}, \citenamefont {Angioletti-Uberti}, \citenamefont {Mognetti},\ and\
  \citenamefont {Frenkel}}]{Varilly_JChemPhys_2012}%
  \BibitemOpen
  \bibfield  {author} {\bibinfo {author} {\bibfnamefont {P.}~\bibnamefont
  {Varilly}}, \bibinfo {author} {\bibfnamefont {S.}~\bibnamefont
  {Angioletti-Uberti}}, \bibinfo {author} {\bibfnamefont {B.~M.}\ \bibnamefont
  {Mognetti}}, \ and\ \bibinfo {author} {\bibfnamefont {D.}~\bibnamefont
  {Frenkel}},\ }\href@noop {} {\bibfield  {journal} {\bibinfo  {journal} {J.
  Chem. Phys.}\ }\textbf {\bibinfo {volume} {137}},\ \bibinfo {pages} {094108}
  (\bibinfo {year} {2012})}\BibitemShut {NoStop}%
\bibitem [{\citenamefont {Angioletti-Uberti}\ \emph {et~al.}(2013)\citenamefont
  {Angioletti-Uberti}, \citenamefont {Varilly}, \citenamefont {Mognetti},
  \citenamefont {Tkachenko},\ and\ \citenamefont
  {Frenkel}}]{Angioletti-Uberti_JCP_2013}%
  \BibitemOpen
  \bibfield  {author} {\bibinfo {author} {\bibfnamefont {S.}~\bibnamefont
  {Angioletti-Uberti}}, \bibinfo {author} {\bibfnamefont {P.}~\bibnamefont
  {Varilly}}, \bibinfo {author} {\bibfnamefont {B.~M.}\ \bibnamefont
  {Mognetti}}, \bibinfo {author} {\bibfnamefont {A.~V.}\ \bibnamefont
  {Tkachenko}}, \ and\ \bibinfo {author} {\bibfnamefont {D.}~\bibnamefont
  {Frenkel}},\ }\href@noop {} {\bibfield  {journal} {\bibinfo  {journal} {J.
  Chem. Phys.}\ }\textbf {\bibinfo {volume} {138}},\ \bibinfo {pages} {021102}
  (\bibinfo {year} {2013})}\BibitemShut {NoStop}%
\bibitem [{\citenamefont {Mognetti}\ \emph
  {et~al.}(2012{\natexlab{a}})\citenamefont {Mognetti}, \citenamefont
  {Leunissen},\ and\ \citenamefont {Frenkel}}]{Mognetti_SoftMatter_2012}%
  \BibitemOpen
  \bibfield  {author} {\bibinfo {author} {\bibfnamefont {B.~M.}\ \bibnamefont
  {Mognetti}}, \bibinfo {author} {\bibfnamefont {M.~E.}\ \bibnamefont
  {Leunissen}}, \ and\ \bibinfo {author} {\bibfnamefont {D.}~\bibnamefont
  {Frenkel}},\ }\href {\doibase 10.1039/C2SM06635A} {\bibfield  {journal}
  {\bibinfo  {journal} {Soft Matter}\ }\textbf {\bibinfo {volume} {8}},\
  \bibinfo {pages} {2213} (\bibinfo {year} {2012}{\natexlab{a}})}\BibitemShut
  {NoStop}%
\bibitem [{\citenamefont {Mognetti}\ \emph
  {et~al.}(2012{\natexlab{b}})\citenamefont {Mognetti}, \citenamefont
  {Varilly}, \citenamefont {Angioletti-Uberti}, \citenamefont
  {Martinez-Veracoechea}, \citenamefont {Dobnikar}, \citenamefont {Leunissen},\
  and\ \citenamefont {Frenkel}}]{Mognetti_PNAS_2012}%
  \BibitemOpen
  \bibfield  {author} {\bibinfo {author} {\bibfnamefont {B.~M.}\ \bibnamefont
  {Mognetti}}, \bibinfo {author} {\bibfnamefont {P.}~\bibnamefont {Varilly}},
  \bibinfo {author} {\bibfnamefont {S.}~\bibnamefont {Angioletti-Uberti}},
  \bibinfo {author} {\bibfnamefont {F.~J.}\ \bibnamefont
  {Martinez-Veracoechea}}, \bibinfo {author} {\bibfnamefont {J.}~\bibnamefont
  {Dobnikar}}, \bibinfo {author} {\bibfnamefont {M.~E.}\ \bibnamefont
  {Leunissen}}, \ and\ \bibinfo {author} {\bibfnamefont {D.}~\bibnamefont
  {Frenkel}},\ }\href {http://www.pnas.org/content/109/7/E378.short} {\bibfield
   {journal} {\bibinfo  {journal} {Proc. Natl. Acad. Sci. U.S.A.}\ }\textbf
  {\bibinfo {volume} {109}},\ \bibinfo {pages} {E378} (\bibinfo {year}
  {2012}{\natexlab{b}})}\BibitemShut {NoStop}%
\bibitem [{\citenamefont {Leunissen}\ and\ \citenamefont
  {Frenkel}(2011)}]{Leunissen_JCP_2011}%
  \BibitemOpen
  \bibfield  {author} {\bibinfo {author} {\bibfnamefont {M.~E.}\ \bibnamefont
  {Leunissen}}\ and\ \bibinfo {author} {\bibfnamefont {D.}~\bibnamefont
  {Frenkel}},\ }\href {\doibase http://dx.doi.org/10.1063/1.3557794} {\bibfield
   {journal} {\bibinfo  {journal} {J. Chem. Phys.}\ }\textbf {\bibinfo {volume}
  {134}},\ \bibinfo {eid} {084702} (\bibinfo {year} {2011})}\BibitemShut
  {NoStop}%
\bibitem [{\citenamefont {Hadorn}\ \emph {et~al.}(2012)\citenamefont {Hadorn},
  \citenamefont {Boenzli}, \citenamefont {S{\o}rensen}, \citenamefont
  {Fellermann}, \citenamefont {Eggenberger~Hotz},\ and\ \citenamefont
  {Hanczyc}}]{Hadorn_PNAS_2012}%
  \BibitemOpen
  \bibfield  {author} {\bibinfo {author} {\bibfnamefont {M.}~\bibnamefont
  {Hadorn}}, \bibinfo {author} {\bibfnamefont {E.}~\bibnamefont {Boenzli}},
  \bibinfo {author} {\bibfnamefont {K.~T.}\ \bibnamefont {S{\o}rensen}},
  \bibinfo {author} {\bibfnamefont {H.}~\bibnamefont {Fellermann}}, \bibinfo
  {author} {\bibfnamefont {P.}~\bibnamefont {Eggenberger~Hotz}}, \ and\
  \bibinfo {author} {\bibfnamefont {M.~M.}\ \bibnamefont {Hanczyc}},\
  }\href@noop {} {\bibfield  {journal} {\bibinfo  {journal} {Proc. Natl. Acad.
  Sci. U.S.A.}\ }\textbf {\bibinfo {volume} {109}},\ \bibinfo {pages} {20320}
  (\bibinfo {year} {2012})}\BibitemShut {NoStop}%
\bibitem [{\citenamefont {Feng}\ \emph {et~al.}(2013)\citenamefont {Feng},
  \citenamefont {Pontani}, \citenamefont {Dreyfus}, \citenamefont {Chaikin},\
  and\ \citenamefont {Brujic}}]{Feng_SoftMatter_2013}%
  \BibitemOpen
  \bibfield  {author} {\bibinfo {author} {\bibfnamefont {L.}~\bibnamefont
  {Feng}}, \bibinfo {author} {\bibfnamefont {L.-L.}\ \bibnamefont {Pontani}},
  \bibinfo {author} {\bibfnamefont {R.}~\bibnamefont {Dreyfus}}, \bibinfo
  {author} {\bibfnamefont {P.}~\bibnamefont {Chaikin}}, \ and\ \bibinfo
  {author} {\bibfnamefont {J.}~\bibnamefont {Brujic}},\ }\href {\doibase
  10.1039/C3SM51586A} {\bibfield  {journal} {\bibinfo  {journal} {Soft Matter}\
  }\textbf {\bibinfo {volume} {9}},\ \bibinfo {pages} {9816} (\bibinfo {year}
  {2013})}\BibitemShut {NoStop}%
\bibitem [{\citenamefont {Hadorn}\ \emph {et~al.}(2013)\citenamefont {Hadorn},
  \citenamefont {Boenzli}, \citenamefont {S{\o}rensen}, \citenamefont
  {De~Lucrezia}, \citenamefont {Hanczyc},\ and\ \citenamefont
  {Yomo}}]{Hadorn_Langmuir_2013}%
  \BibitemOpen
  \bibfield  {author} {\bibinfo {author} {\bibfnamefont {M.}~\bibnamefont
  {Hadorn}}, \bibinfo {author} {\bibfnamefont {E.}~\bibnamefont {Boenzli}},
  \bibinfo {author} {\bibfnamefont {K.~T.}\ \bibnamefont {S{\o}rensen}},
  \bibinfo {author} {\bibfnamefont {D.}~\bibnamefont {De~Lucrezia}}, \bibinfo
  {author} {\bibfnamefont {M.~M.}\ \bibnamefont {Hanczyc}}, \ and\ \bibinfo
  {author} {\bibfnamefont {T.}~\bibnamefont {Yomo}},\ }\href {\doibase
  10.1021/la402621r} {\bibfield  {journal} {\bibinfo  {journal} {Langmuir}\
  }\textbf {\bibinfo {volume} {29}},\ \bibinfo {pages} {15309} (\bibinfo {year}
  {2013})}\BibitemShut {NoStop}%
\bibitem [{\citenamefont {Hadorn}\ and\ \citenamefont
  {Eggenberger~Hotz}(2010)}]{Hadorn_PLOSOne_2010}%
  \BibitemOpen
  \bibfield  {author} {\bibinfo {author} {\bibfnamefont {M.}~\bibnamefont
  {Hadorn}}\ and\ \bibinfo {author} {\bibfnamefont {P.}~\bibnamefont
  {Eggenberger~Hotz}},\ }\href@noop {} {\bibfield  {journal} {\bibinfo
  {journal} {PLoS ONE}\ }\textbf {\bibinfo {volume} {5}},\ \bibinfo {pages}
  {e9886 EP } (\bibinfo {year} {2010})}\BibitemShut {NoStop}%
\bibitem [{\citenamefont {Beales}\ and\ \citenamefont
  {Vanderlick}(2007)}]{Beales_JPCA_2007}%
  \BibitemOpen
  \bibfield  {author} {\bibinfo {author} {\bibfnamefont {P.~A.}\ \bibnamefont
  {Beales}}\ and\ \bibinfo {author} {\bibfnamefont {T.~K.}\ \bibnamefont
  {Vanderlick}},\ }\bibfield  {booktitle} {\emph {\bibinfo {booktitle} {The
  Journal of Physical Chemistry A}},\ }\href {\doibase 10.1021/jp075792z}
  {\bibfield  {journal} {\bibinfo  {journal} {J. Phys. Chem. A}\ }\textbf
  {\bibinfo {volume} {111}},\ \bibinfo {pages} {12372} (\bibinfo {year}
  {2007})}\BibitemShut {NoStop}%
\bibitem [{\citenamefont {Beales}\ \emph {et~al.}(2011)\citenamefont {Beales},
  \citenamefont {Nam},\ and\ \citenamefont
  {Vanderlick}}]{Beales_SoftMatter_2011}%
  \BibitemOpen
  \bibfield  {author} {\bibinfo {author} {\bibfnamefont {P.~A.}\ \bibnamefont
  {Beales}}, \bibinfo {author} {\bibfnamefont {J.}~\bibnamefont {Nam}}, \ and\
  \bibinfo {author} {\bibfnamefont {T.~K.}\ \bibnamefont {Vanderlick}},\
  }\href@noop {} {\bibfield  {journal} {\bibinfo  {journal} {Soft Matter}\
  }\textbf {\bibinfo {volume} {7}},\ \bibinfo {pages} {1747} (\bibinfo {year}
  {2011})}\BibitemShut {NoStop}%
\bibitem [{\citenamefont {Beales}\ and\ \citenamefont
  {Vanderlick}(2014)}]{Beales_ACIS_2014}%
  \BibitemOpen
  \bibfield  {author} {\bibinfo {author} {\bibfnamefont {P.~A.}\ \bibnamefont
  {Beales}}\ and\ \bibinfo {author} {\bibfnamefont {T.~K.}\ \bibnamefont
  {Vanderlick}},\ }\bibfield  {booktitle} {\emph {\bibinfo {booktitle} {Special
  Issue: Helmuth M{\"o}hwald Honorary Issue}},\ }\href@noop {} {\bibfield
  {journal} {\bibinfo  {journal} {Adv. Coll. Int. Sci.}\ }\textbf {\bibinfo
  {volume} {207}},\ \bibinfo {pages} {290} (\bibinfo {year}
  {2014})}\BibitemShut {NoStop}%
\bibitem [{\citenamefont {Parolini}\ \emph {et~al.}(2015)\citenamefont
  {Parolini}, \citenamefont {Mognetti}, \citenamefont {Kotar}, \citenamefont
  {Eiser}, \citenamefont {Cicuta},\ and\ \citenamefont
  {Di~Michele}}]{Parolini_NatComms_2015}%
  \BibitemOpen
  \bibfield  {author} {\bibinfo {author} {\bibfnamefont {L.}~\bibnamefont
  {Parolini}}, \bibinfo {author} {\bibfnamefont {B.~M.}\ \bibnamefont
  {Mognetti}}, \bibinfo {author} {\bibfnamefont {J.}~\bibnamefont {Kotar}},
  \bibinfo {author} {\bibfnamefont {E.}~\bibnamefont {Eiser}}, \bibinfo
  {author} {\bibfnamefont {P.}~\bibnamefont {Cicuta}}, \ and\ \bibinfo {author}
  {\bibfnamefont {L.}~\bibnamefont {Di~Michele}},\ }\href
  {http://dx.doi.org/10.1038/ncomms6948} {\bibfield  {journal} {\bibinfo
  {journal} {Nat Commun}\ }\textbf {\bibinfo {volume} {6}},\ \bibinfo {pages}
  {5948} (\bibinfo {year} {2015})}\BibitemShut {NoStop}%
\bibitem [{\citenamefont {Angioletti-Uberti}\ \emph {et~al.}(2014)\citenamefont
  {Angioletti-Uberti}, \citenamefont {Varilly}, \citenamefont {Mognetti},\ and\
  \citenamefont {Frenkel}}]{Angioletti-Uberti_PRL_2014}%
  \BibitemOpen
  \bibfield  {author} {\bibinfo {author} {\bibfnamefont {S.}~\bibnamefont
  {Angioletti-Uberti}}, \bibinfo {author} {\bibfnamefont {P.}~\bibnamefont
  {Varilly}}, \bibinfo {author} {\bibfnamefont {B.~M.}\ \bibnamefont
  {Mognetti}}, \ and\ \bibinfo {author} {\bibfnamefont {D.}~\bibnamefont
  {Frenkel}},\ }\href {http://link.aps.org/doi/10.1103/PhysRevLett.113.128303}
  {\bibfield  {journal} {\bibinfo  {journal} {Phys. Rev. Lett.}\ }\textbf
  {\bibinfo {volume} {113}},\ \bibinfo {pages} {128303} (\bibinfo {year}
  {2014})}\BibitemShut {NoStop}%
\bibitem [{\citenamefont {Yoon}\ \emph {et~al.}(2009)\citenamefont {Yoon},
  \citenamefont {Hong}, \citenamefont {Brown}, \citenamefont {Kim},
  \citenamefont {Kang}, \citenamefont {Lew},\ and\ \citenamefont
  {Cicuta}}]{Yoon_BJ_2009}%
  \BibitemOpen
  \bibfield  {author} {\bibinfo {author} {\bibfnamefont {Y.-Z.}\ \bibnamefont
  {Yoon}}, \bibinfo {author} {\bibfnamefont {H.}~\bibnamefont {Hong}}, \bibinfo
  {author} {\bibfnamefont {A.}~\bibnamefont {Brown}}, \bibinfo {author}
  {\bibfnamefont {D.~C.}\ \bibnamefont {Kim}}, \bibinfo {author} {\bibfnamefont
  {D.~J.}\ \bibnamefont {Kang}}, \bibinfo {author} {\bibfnamefont {V.~L.}\
  \bibnamefont {Lew}}, \ and\ \bibinfo {author} {\bibfnamefont
  {P.}~\bibnamefont {Cicuta}},\ }\href {\doibase 10.1016/j.bpj.2009.06.028}
  {\bibfield  {journal} {\bibinfo  {journal} {Biophysical Journal}\ }\textbf
  {\bibinfo {volume} {97}},\ \bibinfo {pages} {1606} (\bibinfo {year}
  {2009})}\BibitemShut {NoStop}%
\bibitem [{\citenamefont {HelfrichW.}\ and\ \citenamefont
  {M.}(1984)}]{Helfrich}%
  \BibitemOpen
  \bibfield  {author} {\bibinfo {author} {\bibnamefont {HelfrichW.}}\ and\
  \bibinfo {author} {\bibfnamefont {S.~R.}\ \bibnamefont {M.}},\ }\href@noop {}
  {\bibfield  {journal} {\bibinfo  {journal} {Nuovo Cimento D}\ }\textbf
  {\bibinfo {volume} {3}},\ \bibinfo {pages} {137} (\bibinfo {year}
  {1984})}\BibitemShut {NoStop}%
\bibitem [{\citenamefont {P{\'e}cr{\'e}aux}\ \emph {et~al.}(2004)\citenamefont
  {P{\'e}cr{\'e}aux}, \citenamefont {D{\"o}bereiner}, \citenamefont {Prost},
  \citenamefont {Joanny},\ and\ \citenamefont
  {Bassereau}}]{Pecreaux_EPJE_2004}%
  \BibitemOpen
  \bibfield  {author} {\bibinfo {author} {\bibfnamefont {J.}~\bibnamefont
  {P{\'e}cr{\'e}aux}}, \bibinfo {author} {\bibfnamefont {H.~G.}\ \bibnamefont
  {D{\"o}bereiner}}, \bibinfo {author} {\bibfnamefont {J.}~\bibnamefont
  {Prost}}, \bibinfo {author} {\bibfnamefont {J.~F.}\ \bibnamefont {Joanny}}, \
  and\ \bibinfo {author} {\bibfnamefont {P.}~\bibnamefont {Bassereau}},\
  }\bibfield  {booktitle} {\emph {\bibinfo {booktitle} {The European Physical
  Journal E}},\ }\href {\doibase 10.1140/epje/i2004-10001-9} {\bibfield
  {journal} {\bibinfo  {journal} {Eur. Phys. J. E Soft Matter}\ }\textbf
  {\bibinfo {volume} {13}},\ \bibinfo {pages} {277} (\bibinfo {year}
  {2004})}\BibitemShut {NoStop}%
\bibitem [{\citenamefont {Smith}\ \emph {et~al.}(1996)\citenamefont {Smith},
  \citenamefont {Cui},\ and\ \citenamefont {Bustamante}}]{Smith_Science_1996}%
  \BibitemOpen
  \bibfield  {author} {\bibinfo {author} {\bibfnamefont {S.~B.}\ \bibnamefont
  {Smith}}, \bibinfo {author} {\bibfnamefont {Y.}~\bibnamefont {Cui}}, \ and\
  \bibinfo {author} {\bibfnamefont {C.}~\bibnamefont {Bustamante}},\
  }\href@noop {} {\bibfield  {journal} {\bibinfo  {journal} {Science}\ }\textbf
  {\bibinfo {volume} {271}},\ \bibinfo {pages} {795} (\bibinfo {year}
  {1996})}\BibitemShut {NoStop}%
\bibitem [{\citenamefont {Angelova}\ and\ \citenamefont
  {Dimitrov}(1989)}]{Angelova_FD_1989}%
  \BibitemOpen
  \bibfield  {author} {\bibinfo {author} {\bibfnamefont {M.~I.}\ \bibnamefont
  {Angelova}}\ and\ \bibinfo {author} {\bibfnamefont {D.}~\bibnamefont
  {Dimitrov}},\ }\href@noop {} {\bibfield  {journal} {\bibinfo  {journal}
  {Faraday Discuss. Chem. Soc.}\ }\textbf {\bibinfo {volume} {81}},\ \bibinfo
  {pages} {303} (\bibinfo {year} {1989})}\BibitemShut {NoStop}%
\bibitem [{\citenamefont {Angelova}\ \emph {et~al.}(1992)\citenamefont
  {Angelova}, \citenamefont {Sol\'{e}au}, \citenamefont {M\'{e}l\'{e}ard},
  \citenamefont {Faucon},\ and\ \citenamefont {Bothorel}}]{Angelova_PCPS_1992}%
  \BibitemOpen
  \bibfield  {author} {\bibinfo {author} {\bibfnamefont {M.~I.}\ \bibnamefont
  {Angelova}}, \bibinfo {author} {\bibfnamefont {S.}~\bibnamefont
  {Sol\'{e}au}}, \bibinfo {author} {\bibfnamefont {P.}~\bibnamefont
  {M\'{e}l\'{e}ard}}, \bibinfo {author} {\bibfnamefont {J.-F.}\ \bibnamefont
  {Faucon}}, \ and\ \bibinfo {author} {\bibfnamefont {P.}~\bibnamefont
  {Bothorel}},\ }\href@noop {} {\bibfield  {journal} {\bibinfo  {journal} {Pro.
  Colloid Polym. Sci.}\ }\textbf {\bibinfo {volume} {89}},\ \bibinfo {pages}
  {127} (\bibinfo {year} {1992})}\BibitemShut {NoStop}%
\bibitem [{\citenamefont {Cremer}\ and\ \citenamefont
  {Boxer}(1999)}]{Cremer_JPCB_1999}%
  \BibitemOpen
  \bibfield  {author} {\bibinfo {author} {\bibfnamefont {P.~S.}\ \bibnamefont
  {Cremer}}\ and\ \bibinfo {author} {\bibfnamefont {S.~G.}\ \bibnamefont
  {Boxer}},\ }\bibfield  {booktitle} {\emph {\bibinfo {booktitle} {The Journal
  of Physical Chemistry B}},\ }\href {\doibase 10.1021/jp983996x} {\bibfield
  {journal} {\bibinfo  {journal} {J. Phys. Chem. B}\ }\textbf {\bibinfo
  {volume} {103}},\ \bibinfo {pages} {2554} (\bibinfo {year}
  {1999})}\BibitemShut {NoStop}%
\bibitem [{\citenamefont {Brian}\ and\ \citenamefont
  {McConnell}(1984)}]{Brian_PNAS_1984}%
  \BibitemOpen
  \bibfield  {author} {\bibinfo {author} {\bibfnamefont {A.~A.}\ \bibnamefont
  {Brian}}\ and\ \bibinfo {author} {\bibfnamefont {H.~M.}\ \bibnamefont
  {McConnell}},\ }\href {http://www.pnas.org/content/81/19/6159.abstract}
  {\bibfield  {journal} {\bibinfo  {journal} {Proc. Natl. Acad. Sci. U.S.A.}\
  }\textbf {\bibinfo {volume} {81}},\ \bibinfo {pages} {6159} (\bibinfo {year}
  {1984})}\BibitemShut {NoStop}%
\bibitem [{Note1()}]{Note1}%
  \BibitemOpen
  \bibinfo {note} {Matlab's \protect \textit {fminsearch} function}\BibitemShut
  {NoStop}%
\bibitem [{\citenamefont {Ramachandran}\ \emph {et~al.}(2010)\citenamefont
  {Ramachandran}, \citenamefont {Anderson}, \citenamefont {Leal},\ and\
  \citenamefont {Israelachvili}}]{Ramachandran_Langmuir_2010}%
  \BibitemOpen
  \bibfield  {author} {\bibinfo {author} {\bibfnamefont {A.}~\bibnamefont
  {Ramachandran}}, \bibinfo {author} {\bibfnamefont {T.~H.}\ \bibnamefont
  {Anderson}}, \bibinfo {author} {\bibfnamefont {L.~G.}\ \bibnamefont {Leal}},
  \ and\ \bibinfo {author} {\bibfnamefont {J.~N.}\ \bibnamefont
  {Israelachvili}},\ }\href {\doibase 10.1021/la1023168} {\bibfield  {journal}
  {\bibinfo  {journal} {Langmuir}\ }\textbf {\bibinfo {volume} {27}},\ \bibinfo
  {pages} {59} (\bibinfo {year} {2010})}\BibitemShut {NoStop}%
\bibitem [{\citenamefont {Evans}\ and\ \citenamefont
  {Needham}(1987)}]{Evans_JPC_1987}%
  \BibitemOpen
  \bibfield  {author} {\bibinfo {author} {\bibfnamefont {E.}~\bibnamefont
  {Evans}}\ and\ \bibinfo {author} {\bibfnamefont {D.}~\bibnamefont
  {Needham}},\ }\href@noop {} {\bibfield  {journal} {\bibinfo  {journal} {J.
  Phys. Chem.}\ }\textbf {\bibinfo {volume} {91}},\ \bibinfo {pages} {4219}
  (\bibinfo {year} {1987})}\BibitemShut {NoStop}%
\bibitem [{\citenamefont {Bracha}\ \emph {et~al.}(2013)\citenamefont {Bracha},
  \citenamefont {Karzbrun}, \citenamefont {Shemer}, \citenamefont {Pincus},\
  and\ \citenamefont {Bar-Ziv}}]{Bracha_PNAS_2013}%
  \BibitemOpen
  \bibfield  {author} {\bibinfo {author} {\bibfnamefont {D.}~\bibnamefont
  {Bracha}}, \bibinfo {author} {\bibfnamefont {E.}~\bibnamefont {Karzbrun}},
  \bibinfo {author} {\bibfnamefont {G.}~\bibnamefont {Shemer}}, \bibinfo
  {author} {\bibfnamefont {P.~A.}\ \bibnamefont {Pincus}}, \ and\ \bibinfo
  {author} {\bibfnamefont {R.~H.}\ \bibnamefont {Bar-Ziv}},\ }\href {\doibase
  10.1073/pnas.1220076110} {\bibfield  {journal} {\bibinfo  {journal} {Proc.
  Natl. Acad. Sci. U.S.A.}\ }\textbf {\bibinfo {volume} {110}},\ \bibinfo
  {pages} {4534} (\bibinfo {year} {2013})}\BibitemShut {NoStop}%
\bibitem [{\citenamefont {SantaLucia}(1998)}]{SantaLucia_PNAS_1998}%
  \BibitemOpen
  \bibfield  {author} {\bibinfo {author} {\bibfnamefont {J.}~\bibnamefont
  {SantaLucia}},\ }\href@noop {} {\bibfield  {journal} {\bibinfo  {journal}
  {Proc. Natl. Acad. Sci. U.S.A.}\ }\textbf {\bibinfo {volume} {95}},\ \bibinfo
  {pages} {1460} (\bibinfo {year} {1998})}\BibitemShut {NoStop}%
\bibitem [{\citenamefont {Di~Michele}\ \emph
  {et~al.}(2014{\natexlab{b}})\citenamefont {Di~Michele}, \citenamefont
  {Mognetti}, \citenamefont {Yanagishima}, \citenamefont {Varilly},
  \citenamefont {Ruff}, \citenamefont {Frenkel},\ and\ \citenamefont
  {Eiser}}]{Di-Michele_JACS_2014}%
  \BibitemOpen
  \bibfield  {author} {\bibinfo {author} {\bibfnamefont {L.}~\bibnamefont
  {Di~Michele}}, \bibinfo {author} {\bibfnamefont {B.~M.}\ \bibnamefont
  {Mognetti}}, \bibinfo {author} {\bibfnamefont {T.}~\bibnamefont
  {Yanagishima}}, \bibinfo {author} {\bibfnamefont {P.}~\bibnamefont
  {Varilly}}, \bibinfo {author} {\bibfnamefont {Z.}~\bibnamefont {Ruff}},
  \bibinfo {author} {\bibfnamefont {D.}~\bibnamefont {Frenkel}}, \ and\
  \bibinfo {author} {\bibfnamefont {E.}~\bibnamefont {Eiser}},\ }\href
  {\doibase 10.1021/ja500027v} {\bibfield  {journal} {\bibinfo  {journal} {J.
  Am. Chem. Soc.}\ }\textbf {\bibinfo {volume} {136}},\ \bibinfo {pages} {6538}
  (\bibinfo {year} {2014}{\natexlab{b}})}\BibitemShut {NoStop}%
\bibitem [{\citenamefont {Angioletti-Uberti}\ \emph {et~al.}(2012)\citenamefont
  {Angioletti-Uberti}, \citenamefont {Mognetti},\ and\ \citenamefont
  {Frenkel}}]{Angioletti-Uberti_NMat_2012}%
  \BibitemOpen
  \bibfield  {author} {\bibinfo {author} {\bibfnamefont {S.}~\bibnamefont
  {Angioletti-Uberti}}, \bibinfo {author} {\bibfnamefont {B.~M.}\ \bibnamefont
  {Mognetti}}, \ and\ \bibinfo {author} {\bibfnamefont {D.}~\bibnamefont
  {Frenkel}},\ }\href@noop {} {\bibfield  {journal} {\bibinfo  {journal} {Nat.
  Mater.}\ }\textbf {\bibinfo {volume} {11}},\ \bibinfo {pages} {518} (\bibinfo
  {year} {2012})}\BibitemShut {NoStop}%
\bibitem [{\citenamefont {Hu}\ \emph {et~al.}(2013)\citenamefont {Hu},
  \citenamefont {Lipowsky},\ and\ \citenamefont {Weikl}}]{Hu_PNAS_2013}%
  \BibitemOpen
  \bibfield  {author} {\bibinfo {author} {\bibfnamefont {J.}~\bibnamefont
  {Hu}}, \bibinfo {author} {\bibfnamefont {R.}~\bibnamefont {Lipowsky}}, \ and\
  \bibinfo {author} {\bibfnamefont {T.~R.}\ \bibnamefont {Weikl}},\ }\href
  {http://www.pnas.org/content/110/38/15283.abstract N2 - Cell adhesion and the
  adhesion of vesicles to the membranes of cells or organelles are pivotal for
  immune responses, tissue formation, and cell signaling. The adhesion
  processes depend sensitively on the binding constant of the membrane-anchored
  receptor and ligand proteins that mediate adhesion, but this constant is
  difficult to measure in experiments. We have investigated the binding of
  membrane-anchored receptor and ligand proteins with molecular dynamics
  simulations. We find that the binding constant of the anchored proteins
  strongly decreases with the membrane roughness caused by thermally excited
  membrane shape fluctuations on nanoscales. We present a theory that explains
  the roughness dependence of the binding constant for the anchored proteins
  from membrane confinement and that relates this constant to the binding
  constant of soluble proteins without membrane anchors. Because the binding
  constant of soluble proteins is readily accessible in experiments, our
  results provide a useful route to compute the binding constant of
  membrane-anchored receptor and ligand proteins.} {\bibfield  {journal}
  {\bibinfo  {journal} {Proceedings of the National Academy of Sciences}\
  }\textbf {\bibinfo {volume} {110}},\ \bibinfo {pages} {15283} (\bibinfo
  {year} {2013})}\BibitemShut {NoStop}%
\bibitem [{\citenamefont {Rawicz}\ \emph {et~al.}(2000)\citenamefont {Rawicz},
  \citenamefont {Olbrich}, \citenamefont {McIntosh}, \citenamefont {Needham},\
  and\ \citenamefont {Evans}}]{Rawicz_BJ_2000}%
  \BibitemOpen
  \bibfield  {author} {\bibinfo {author} {\bibfnamefont {W.}~\bibnamefont
  {Rawicz}}, \bibinfo {author} {\bibfnamefont {K.~C.}\ \bibnamefont {Olbrich}},
  \bibinfo {author} {\bibfnamefont {T.}~\bibnamefont {McIntosh}}, \bibinfo
  {author} {\bibfnamefont {D.}~\bibnamefont {Needham}}, \ and\ \bibinfo
  {author} {\bibfnamefont {E.}~\bibnamefont {Evans}},\ }\href@noop {}
  {\bibfield  {journal} {\bibinfo  {journal} {Biophys. J.}\ }\textbf {\bibinfo
  {volume} {79}},\ \bibinfo {pages} {328} (\bibinfo {year} {2000})}\BibitemShut
  {NoStop}%
\bibitem [{\citenamefont {Rawicz}\ \emph {et~al.}(2008)\citenamefont {Rawicz},
  \citenamefont {Smith}, \citenamefont {McIntosh}, \citenamefont {Simon},\ and\
  \citenamefont {Evans}}]{Rawicz_BJ_2008}%
  \BibitemOpen
  \bibfield  {author} {\bibinfo {author} {\bibfnamefont {W.}~\bibnamefont
  {Rawicz}}, \bibinfo {author} {\bibfnamefont {B.~A.}\ \bibnamefont {Smith}},
  \bibinfo {author} {\bibfnamefont {T.~J.}\ \bibnamefont {McIntosh}}, \bibinfo
  {author} {\bibfnamefont {S.~A.}\ \bibnamefont {Simon}}, \ and\ \bibinfo
  {author} {\bibfnamefont {E.}~\bibnamefont {Evans}},\ }\href@noop {}
  {\bibfield  {journal} {\bibinfo  {journal} {Biophys. J.}\ }\textbf {\bibinfo
  {volume} {94}},\ \bibinfo {pages} {4725} (\bibinfo {year}
  {2008})}\BibitemShut {NoStop}%
\bibitem [{\citenamefont {Fa}\ \emph {et~al.}(2007)\citenamefont {Fa},
  \citenamefont {Lins}, \citenamefont {Courtoy}, \citenamefont {Dufr{\^e}ne},
  \citenamefont {Van Der~Smissen}, \citenamefont {Brasseur}, \citenamefont
  {Tyteca},\ and\ \citenamefont {Mingeot-Leclercq}}]{Fa_BBA_2007}%
  \BibitemOpen
  \bibfield  {author} {\bibinfo {author} {\bibfnamefont {N.}~\bibnamefont
  {Fa}}, \bibinfo {author} {\bibfnamefont {L.}~\bibnamefont {Lins}}, \bibinfo
  {author} {\bibfnamefont {P.~J.}\ \bibnamefont {Courtoy}}, \bibinfo {author}
  {\bibfnamefont {Y.}~\bibnamefont {Dufr{\^e}ne}}, \bibinfo {author}
  {\bibfnamefont {P.}~\bibnamefont {Van Der~Smissen}}, \bibinfo {author}
  {\bibfnamefont {R.}~\bibnamefont {Brasseur}}, \bibinfo {author}
  {\bibfnamefont {D.}~\bibnamefont {Tyteca}}, \ and\ \bibinfo {author}
  {\bibfnamefont {M.~P.}\ \bibnamefont {Mingeot-Leclercq}},\ }\href@noop {}
  {\bibfield  {journal} {\bibinfo  {journal} {BBA - Biomembranes}\ }\textbf
  {\bibinfo {volume} {1768}},\ \bibinfo {pages} {1830} (\bibinfo {year}
  {2007})}\BibitemShut {NoStop}%
\bibitem [{\citenamefont {Pan}\ \emph {et~al.}(2008)\citenamefont {Pan},
  \citenamefont {Tristram-Nagle}, \citenamefont {Ku{\v c}erka},\ and\
  \citenamefont {Nagle}}]{Pan_BJ_2008}%
  \BibitemOpen
  \bibfield  {author} {\bibinfo {author} {\bibfnamefont {J.}~\bibnamefont
  {Pan}}, \bibinfo {author} {\bibfnamefont {S.}~\bibnamefont {Tristram-Nagle}},
  \bibinfo {author} {\bibfnamefont {N.}~\bibnamefont {Ku{\v c}erka}}, \ and\
  \bibinfo {author} {\bibfnamefont {J.~F.}\ \bibnamefont {Nagle}},\ }\href@noop
  {} {\bibfield  {journal} {\bibinfo  {journal} {Biophys. J.}\ }\textbf
  {\bibinfo {volume} {94}},\ \bibinfo {pages} {117} (\bibinfo {year}
  {2008})}\BibitemShut {NoStop}%
\bibitem [{\citenamefont {Stengel}\ \emph {et~al.}(2008)\citenamefont
  {Stengel}, \citenamefont {Simonsson}, \citenamefont {Campbell},\ and\
  \citenamefont {H{\"o}{\"o}k}}]{Stengel_JCPB_2008}%
  \BibitemOpen
  \bibfield  {author} {\bibinfo {author} {\bibfnamefont {G.}~\bibnamefont
  {Stengel}}, \bibinfo {author} {\bibfnamefont {L.}~\bibnamefont {Simonsson}},
  \bibinfo {author} {\bibfnamefont {R.~A.}\ \bibnamefont {Campbell}}, \ and\
  \bibinfo {author} {\bibfnamefont {F.}~\bibnamefont {H{\"o}{\"o}k}},\
  }\bibfield  {booktitle} {\emph {\bibinfo {booktitle} {The Journal of Physical
  Chemistry B}},\ }\href {\doibase 10.1021/jp802005b} {\bibfield  {journal}
  {\bibinfo  {journal} {The Journal of Physical Chemistry B}\ }\textbf
  {\bibinfo {volume} {112}},\ \bibinfo {pages} {8264} (\bibinfo {year}
  {2008})}\BibitemShut {NoStop}%
\bibitem [{\citenamefont {Yuan}\ \emph {et~al.}(2007)\citenamefont {Yuan},
  \citenamefont {Griffin}, \citenamefont {Phelps}, \citenamefont {Buschmann},
  \citenamefont {Weston},\ and\ \citenamefont {Greenbaum}}]{Yuan_NAR_2007}%
  \BibitemOpen
  \bibfield  {author} {\bibinfo {author} {\bibfnamefont {F.}~\bibnamefont
  {Yuan}}, \bibinfo {author} {\bibfnamefont {L.}~\bibnamefont {Griffin}},
  \bibinfo {author} {\bibfnamefont {L.}~\bibnamefont {Phelps}}, \bibinfo
  {author} {\bibfnamefont {V.}~\bibnamefont {Buschmann}}, \bibinfo {author}
  {\bibfnamefont {K.}~\bibnamefont {Weston}}, \ and\ \bibinfo {author}
  {\bibfnamefont {N.~L.}\ \bibnamefont {Greenbaum}},\ }\href
  {http://nar.oxfordjournals.org/content/35/9/2833.abstract N2 - U2 and U6
  snRNAs pair to form a phylogenetically conserved complex at the catalytic
  core of the spliceosome. Interactions with divalent metal ions, particularly
  Mg(II), at specific sites are essential for its folding and catalytic
  activity. We used a novel F\UTF{00F6}rster resonance energy transfer (FRET)
  method between site-bound luminescent lanthanide ions and a covalently
  attached fluorescent dye, combined with supporting stoichiometric and
  mutational studies, to determine locations of site-bound Tb(III) within the
  human U2\UTF{2013}U6 complex. At pH 7.2, we detected three metal-ion-binding
  sites in: (1) the consensus ACACAGA sequence, which forms the internal loop
  between helices I and III; (2) the four-way junction, which contains the
  conserved AGC triad; and (3) the internal loop of the U6 intra-molecular stem
  loop (ISL). Binding at each of these sites is supported by previous
  phosphorothioate substitution studies and, in the case of the ISL site, by
  NMR. Binding of Tb(III) at the four-way junction and the ISL sites was found
  to be pH-dependent, with no ion binding observed below pH 6 and 7,
  respectively. This pH dependence of metal ion binding suggests that the local
  environment may play a role in the binding of metal ions, which may impact on
  splicing activity.} {\bibfield  {journal} {\bibinfo  {journal} {Nucleic Acids
  Research}\ }\textbf {\bibinfo {volume} {35}},\ \bibinfo {pages} {2833}
  (\bibinfo {year} {2007})}\BibitemShut {NoStop}%
\bibitem [{\citenamefont {McCann}\ \emph {et~al.}(2010)\citenamefont {McCann},
  \citenamefont {Choi}, \citenamefont {Zheng}, \citenamefont {Weninger},\ and\
  \citenamefont {Bowen}}]{McCann_BJ_2010}%
  \BibitemOpen
  \bibfield  {author} {\bibinfo {author} {\bibfnamefont {J.~J.}\ \bibnamefont
  {McCann}}, \bibinfo {author} {\bibfnamefont {U.~B.}\ \bibnamefont {Choi}},
  \bibinfo {author} {\bibfnamefont {L.}~\bibnamefont {Zheng}}, \bibinfo
  {author} {\bibfnamefont {K.}~\bibnamefont {Weninger}}, \ and\ \bibinfo
  {author} {\bibfnamefont {M.~E.}\ \bibnamefont {Bowen}},\ }\href {\doibase
  http://dx.doi.org/10.1016/j.bpj.2010.04.063} {\bibfield  {journal} {\bibinfo
  {journal} {Biophys. J.}\ }\textbf {\bibinfo {volume} {99}},\ \bibinfo {pages}
  {961} (\bibinfo {year} {2010})}\BibitemShut {NoStop}%
\bibitem [{\citenamefont {van~der Meulen}\ and\ \citenamefont
  {Leunissen}(2013)}]{Meulen_JACS_2013}%
  \BibitemOpen
  \bibfield  {author} {\bibinfo {author} {\bibfnamefont {S.~A.~J.}\
  \bibnamefont {van~der Meulen}}\ and\ \bibinfo {author} {\bibfnamefont
  {M.~E.}\ \bibnamefont {Leunissen}},\ }\bibfield  {booktitle} {\emph {\bibinfo
  {booktitle} {Journal of the American Chemical Society}},\ }\href@noop {}
  {\bibfield  {journal} {\bibinfo  {journal} {J. Am. Chem. Soc.}\ }\textbf
  {\bibinfo {volume} {135}},\ \bibinfo {pages} {15129} (\bibinfo {year}
  {2013})}\BibitemShut {NoStop}%
\bibitem [{\citenamefont {Moreira}\ \emph {et~al.}(2005)\citenamefont
  {Moreira}, \citenamefont {You}, \citenamefont {Behlke},\ and\ \citenamefont
  {Owczarzy}}]{Moreira_BBRC_2005}%
  \BibitemOpen
  \bibfield  {author} {\bibinfo {author} {\bibfnamefont {B.~G.}\ \bibnamefont
  {Moreira}}, \bibinfo {author} {\bibfnamefont {Y.}~\bibnamefont {You}},
  \bibinfo {author} {\bibfnamefont {M.~A.}\ \bibnamefont {Behlke}}, \ and\
  \bibinfo {author} {\bibfnamefont {R.}~\bibnamefont {Owczarzy}},\ }\href
  {\doibase http://dx.doi.org/10.1016/j.bbrc.2004.12.035} {\bibfield  {journal}
  {\bibinfo  {journal} {Biochem. Biophys. Res. Commun.}\ }\textbf {\bibinfo
  {volume} {327}},\ \bibinfo {pages} {473} (\bibinfo {year}
  {2005})}\BibitemShut {NoStop}%
\bibitem [{\citenamefont {Dalva}\ \emph {et~al.}(2007)\citenamefont {Dalva},
  \citenamefont {McClelland},\ and\ \citenamefont
  {Kayser}}]{Dalva_NatureRev_2007}%
  \BibitemOpen
  \bibfield  {author} {\bibinfo {author} {\bibfnamefont {M.~B.}\ \bibnamefont
  {Dalva}}, \bibinfo {author} {\bibfnamefont {A.~C.}\ \bibnamefont
  {McClelland}}, \ and\ \bibinfo {author} {\bibfnamefont {M.~S.}\ \bibnamefont
  {Kayser}},\ }\href {http://dx.doi.org/10.1038/nrn2075} {\bibfield  {journal}
  {\bibinfo  {journal} {Nat. Rev. Neurosci.}\ }\textbf {\bibinfo {volume}
  {8}},\ \bibinfo {pages} {206} (\bibinfo {year} {2007})}\BibitemShut {NoStop}%
\bibitem [{\citenamefont {Parsons}\ \emph {et~al.}(2010)\citenamefont
  {Parsons}, \citenamefont {Horwitz},\ and\ \citenamefont
  {Schwartz}}]{Parsons_NatureRev_2010}%
  \BibitemOpen
  \bibfield  {author} {\bibinfo {author} {\bibfnamefont {J.~T.}\ \bibnamefont
  {Parsons}}, \bibinfo {author} {\bibfnamefont {A.~R.}\ \bibnamefont
  {Horwitz}}, \ and\ \bibinfo {author} {\bibfnamefont {M.~A.}\ \bibnamefont
  {Schwartz}},\ }\href {http://dx.doi.org/10.1038/nrm2957} {\bibfield
  {journal} {\bibinfo  {journal} {Nat Rev Mol Cell Biol}\ }\textbf {\bibinfo
  {volume} {11}},\ \bibinfo {pages} {633} (\bibinfo {year} {2010})}\BibitemShut
  {NoStop}%
\bibitem [{\citenamefont {Albets}\ \emph {et~al.}(2007)\citenamefont {Albets},
  \citenamefont {Johnson}, \citenamefont {Lewis}, \citenamefont {Raff},
  \citenamefont {Roberts},\ and\ \citenamefont {Walter}}]{Alberts}%
  \BibitemOpen
  \bibfield  {author} {\bibinfo {author} {\bibfnamefont {B.}~\bibnamefont
  {Albets}}, \bibinfo {author} {\bibfnamefont {A.}~\bibnamefont {Johnson}},
  \bibinfo {author} {\bibfnamefont {J.}~\bibnamefont {Lewis}}, \bibinfo
  {author} {\bibfnamefont {M.}~\bibnamefont {Raff}}, \bibinfo {author}
  {\bibfnamefont {K.}~\bibnamefont {Roberts}}, \ and\ \bibinfo {author}
  {\bibfnamefont {P.}~\bibnamefont {Walter}},\ }\href@noop {} {\emph {\bibinfo
  {title} {Molecular Biology of the Cell}}},\ \bibinfo {edition} {5th}\ ed.\
  (\bibinfo  {publisher} {Garland Science},\ \bibinfo {address} {New York},\
  \bibinfo {year} {2007})\BibitemShut {NoStop}%
\bibitem [{\citenamefont {Bao}\ and\ \citenamefont
  {Suresh}(2003)}]{Bao_NMat_2003}%
  \BibitemOpen
  \bibfield  {author} {\bibinfo {author} {\bibfnamefont {G.}~\bibnamefont
  {Bao}}\ and\ \bibinfo {author} {\bibfnamefont {S.}~\bibnamefont {Suresh}},\
  }\href {http://dx.doi.org/10.1038/nmat1001} {\bibfield  {journal} {\bibinfo
  {journal} {Nat. Mater.}\ }\textbf {\bibinfo {volume} {2}},\ \bibinfo {pages}
  {715} (\bibinfo {year} {2003})}\BibitemShut {NoStop}%
\bibitem [{\citenamefont {Discher}\ \emph {et~al.}(2005)\citenamefont
  {Discher}, \citenamefont {Janmey},\ and\ \citenamefont
  {Wang}}]{Discher_Science_2005}%
  \BibitemOpen
  \bibfield  {author} {\bibinfo {author} {\bibfnamefont {D.~E.}\ \bibnamefont
  {Discher}}, \bibinfo {author} {\bibfnamefont {P.}~\bibnamefont {Janmey}}, \
  and\ \bibinfo {author} {\bibfnamefont {Y.-l.}\ \bibnamefont {Wang}},\ }\href
  {http://www.sciencemag.org/content/310/5751/1139.abstract N2 - Normal tissue
  cells are generally not viable when suspended in a fluid and are therefore
  said to be anchorage dependent. Such cells must adhere to a solid, but a
  solid can be as rigid as glass or softer than a baby's skin. The behavior of
  some cells on soft materials is characteristic of important phenotypes; for
  example, cell growth on soft agar gels is used to identify cancer cells.
  However, an understanding of how tissue cells—including fibroblasts,
  myocytes, neurons, and other cell types—sense matrix stiffness is just
  emerging with quantitative studies of cells adhering to gels (or to other
  cells) with which elasticity can be tuned to approximate that of tissues. Key
  roles in molecular pathways are played by adhesion complexes and the
  actinmyosin cytoskeleton, whose contractile forces are transmitted through
  transcellular structures. The feedback of local matrix stiffness on cell
  state likely has important implications for development, differentiation,
  disease, and regeneration.} {\bibfield  {journal} {\bibinfo  {journal}
  {Science}\ }\textbf {\bibinfo {volume} {310}},\ \bibinfo {pages} {1139}
  (\bibinfo {year} {2005})}\BibitemShut {NoStop}%
\bibitem [{\citenamefont {Tordeux}\ \emph {et~al.}(2002)\citenamefont
  {Tordeux}, \citenamefont {Fournier},\ and\ \citenamefont
  {Galatola}}]{Tordeux_PRE_2002}%
  \BibitemOpen
  \bibfield  {author} {\bibinfo {author} {\bibfnamefont {C.}~\bibnamefont
  {Tordeux}}, \bibinfo {author} {\bibfnamefont {J.~B.}\ \bibnamefont
  {Fournier}}, \ and\ \bibinfo {author} {\bibfnamefont {P.}~\bibnamefont
  {Galatola}},\ }\href@noop {} {\bibfield  {journal} {\bibinfo  {journal}
  {Phys. Rev. E}\ }\textbf {\bibinfo {volume} {65}},\ \bibinfo {pages} {041912}
  (\bibinfo {year} {2002})}\BibitemShut {NoStop}%
\end{thebibliography}%

\end{document}